\documentclass[amsmath,amssymb,aps,twocolumn,prd,freq,floatfix]{revtex4-2}
\usepackage{graphicx}% Include figure files
\usepackage{dcolumn}% Align table columns on decimal point
\usepackage{bm}% bold math
\usepackage{hyperref}% add hypertext capabilities
\usepackage{color}
\usepackage{subfigure}
\usepackage{flexisym}

\begin{document}
\title{Ultra wideband axion search using a Faraday haloscope}
%%\title{Axion search between $200~\mu$eV and  $10~$meV using the spin light interaction}
%\title{Wideband search of axions using spin light interaction}
%%\title{Detection of dark matter using the magneto optical interaction}
%%\thanks{A footnote to the article title}%

\author{A. T. M. Anishur Rahman}
\altaffiliation{Department of Physics,University of Warwick, Coventry, UK}
\email{anishur.rahman@warwick.ac.uk}
%  \email{Second.Author@institution.edu}

%%ly specified

\begin{abstract}
Dark matter is a major constituent of our universe and the axion is a prime candidate. In this article, it is shown that by exploiting the axion induced magnetization in a magnetic rod and the Faraday effect, axions in the mass range $500$ to $5000~\mu$eV, a part of which ($> 3500~\mu$eV) is currently inaccessible to experiments, can be searched for using the same experimental setup in a year using the existing technologies. The magnetic rod is placed inside a high finesse optical cavity, which by confining the probe light inside it increases the interaction time and thus enhances the Faraday effect. This rotates the plane of polarization of the probe light sufficiently and produces a robust signal. Axions of different mass are selected using a dc magnetic field. Detection is carried out by counting photons in the optical domain using a readily available and high quantum efficiency ($\approx 1$) photon counter in a noise-free environment. An optical interferometric scheme that could provide spectroscopic information about the axion to be searched is also proposed.
\end{abstract}

%% - a device known for its sensitivity to weak signals by $1/(1-R_c)$ times, where $R_c$ is the reflectivity of the cavity mirror

\maketitle

%%need to give reasons why the ability to search a wide mass range is important.  The axion also fulfils the criteria of dark matter \cite{PRESKILL1983127,DINE1983137,ABBOTT1983133} - a major fraction of our known universe.

%%abc \cite{chatzidrosos2021fiberized}

The strong interactions, in violation to the standard model of particle physics, conserve parity ($P$) and the product $CP$ of charge ($C$) and parity \cite{SikivieRMP2021}. To solve these inconsistencies, Peccei and Quinn proposed a scalar field and a pseudo particle called the axion \cite{PecceiQuinn1977}. Considering the Planck scale the lower bound of axion mass has been found to be $10^{-13}~$eV while the upper bound of $10^{-2}~$eV has been determined from the consideration of stellar evolution \cite{SikivieRMP2021}. Cold axions produced via the vacuum realignment process are good dark matter candidates \cite{PRESKILL1983127,DINE1983137,ABBOTT1983133} and may be the main constituent of the dark matter at $10^{-5}~$eV  \cite{SikivieRMP2021,BARBIERI2017135}. However, there are substantial uncertainties - approximately four orders of magnitude, arising from the uncertainty in the composition of the dark matter, their densities and the temperature \cite{SikivieRMP2021}. Significant uncertainties also exist in the velocity and the equilibrium state of the galactic halo \cite{SikivieRMP2021} - the main axionic source of most proposed and current experiments \cite{AdmxPRL2021,DuPRL2018, AsztalozPRL2010, AlesiniPRD2021,BARBIERI2017135,FLOWER2019100306, AravanitakiPRL2014, MadmaxPRL2017}. These theoretical uncertainties mean that experimental searches for detecting such particles need to be wideband \cite{SikivieRMP2021,BARBIERI2017135,AdmxPRL2021}.

%% Axions may have been produced in the early universe in two different states - hot and cold \cite{SikivieRMP2021}. In one model, axions are in thermal equilibrium and their velocity is Maxwell-Boltzmann distributed with a dispersion of $270~$km/s. A contrasting model shows that axions cannot be in thermal equilibrium via their gravitational interaction in a time scale of the age of the universe and their velocity dispersion is only $71~$m/s.
%%Particularly,  These uncertainties mean that the experimental search for the dark matter needs to be as wide as possible.

%% This experiment can search axions Their ability search axions of different mass is limited by the  The ADMX collaboration can search axions mass approximately $50\%$ below and above the resonance frequency of the lowest order TM mode of their cavity \cite{AdmxPRL2021,SikivieRMP2021}.

%%There are many proposed and realized axion search experiments \cite{AlesiniPRD2021,AdmxPRL2021, AravanitakiPRL2014,MadmaxPRL2017,FLOWER2019100306}. For a comprehensive reveiw see \cite{SikivieRMP2021}. 
%Current experimental The ADMX experiment and its variants  Such experiments are suitable for searching axions of masses of few tens of $\mu$eV \cite{SikivieRMP2021} - limited by size of the cavity. In this experiment the axion to be searched for is determined by the frequency of the lowest order TM mode of the microwave cavity.

Major experimental efforts center around converting axions into microwave photons inside a microwave cavity in a strong magnetic field \cite{AdmxPRL2021,DuPRL2018,AsztalozPRL2010}. Over the decades such experiments have ruled out axions between $2.7$ to $4.2~\mu$eV and can search axions of masses up to few tens of $\mu$eV \cite{AdmxPRL2021,DuPRL2018,AsztalozPRL2010,SikivieRMP2021}. Another class of experiments exploits the axion-electron interaction in a ferromagnetic material placed inside a microwave cavity \cite{AlesiniPRD2021,BARBIERI2017135,BARBIERI1989357,FLOWER2019100306} and converts axions into microwave photons. Due to the resonant nature of the microwave cavity based schemes \cite{AdmxPRL2021,DuPRL2018, AsztalozPRL2010,CresciniPRL2020,FLOWER2019100306}, searching axions of substantially different mass than the resonance frequencies of such cavities requires a new cavity implying a prolonged time (decades) for scanning the axion spectrum. Moreover, in most proposed and current axion experiments \cite{AdmxPRL2021, DuPRL2018, AsztalozPRL2010, AlesiniPRD2021,FLOWER2019100306,MadmaxPRL2017}, detection is accomplished by measuring the microwave power produced in the axion to microwave conversion process. Such a conversion yields an extremely small amount of power ($<10^{-20}~$W) - requiring multistage amplifications which are often plagued by the dominant electronic noise \cite{AdmxPRL2021,CresciniPRL2020,FLOWER2019100306,AsztalozPRL2010}. Here, more sensitive detectors such as photon counters are highly desirable \cite{CresciniPRL2020,SikivieRMP2021,FLOWER2019100306,Lamoreaux2013} but are not readily available in the microwave domain.

In this article, we propose an axion search scheme in which the axion-electron interaction initiates oscillation in magnetization in a magnetic rod and a linearly polarized probe light travelling through the sample interacts with this oscillatory magnetization via the Faraday effect and converts microwave frequency axions into optical photons. The magnetic rod is placed inside a high finesse optical cavity which by trapping the probe light for a prolonged period of time greatly increases the interaction time and thus enhances the Faraday effect. On exiting the optical cavity, the plane of polarization of the probe light gets rotated and this produces a signal proportional to the light field power. The signal in the optical domain means that readily available photon counters can be used. The axion to be searched for is selected by adjusting the dc bias magnetic field. Importantly, since the Faraday effect only changes the plane of polarization of the light field, the optical cavity remains on resonance when the magnetic field is changed for searching axions of different mass. This means that axions in the mass range $500~\mu$eV to $5000~\mu$eV can be searched for using the same experimental setup in the time scale of a year. Axion mass $> 3500~\mu$eV is currently inaccessible to experiments.

\begin{figure}
    \centering
    \includegraphics[width=8.5cm]{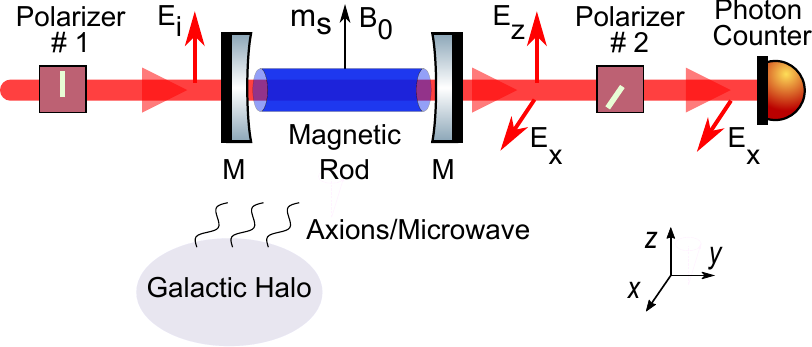}
    \caption{A schematic diagram of the proposed experimental setup. A magnetic rod biased to the saturation using a dc magnetic $B_0$ applied along the $z~$axis is placed inside a high finesse optical cavity. The cavity is formed by the two mirrors (M). Axions from the galactic halos acting as an electromagnetic field and arriving along the $x~$axis couples to the electronic spin in the rod. This initiates precession of the spins or a magnetization and a component of magnetization appears in the $x-y~$plane i.e. $m_y$. Linearly polarized ($z~$axis) light propagating along the $y~$axis interacts with the induced magnetization $m_y$ via the Faraday effect and on exiting the cavity its plane of polarization gets rotated. Polarizer$~\#~ 2$ picks up the $x-$component of the light field. The axes of polarizers $1~\&~ 2$ are orthogonal to each other. The $x-$component of the light field is the axionic signal and detected using a photon counter (see text for more details).}
    \label{fig1}
\end{figure}

%% $a(x)=\sqrt{\frac{n_a\hbar^3}{m_ac}}\exp{[i(cp_0t-\mathbf{p_x}.\mathbf{x})/\hbar]}$ is the axion field with $m_a$ and $n_a$ are the mass and the density of of axion,  $\mathbf{p_x}$ is the momentum axion, $\hbar$ is the reduced Planck constant and $c$ is the speed of light in vacuum.

In the non-relativistic limit, the axion-electron interaction is given by \cite{BARBIERI2017135,SikivieRMP2021}

\begin{eqnarray}
H_{ae}=\frac{g_{ae}}{f_a} \nabla a \cdot\mathbf{S},
\label{eqn0}
\end{eqnarray}

where $\mathbf{S}$ is the spin of the magnetic system, $f_a$ is the axion decay constant and $a$ is the scalar axionic field. The axion-electron coupling strength is given by $g_{ae}$ and in the Dine-Fischler-Srednicki-Zhitnisky (DFSZ) model \cite{DINE1981199,Zhitnitskii1980} is proportional to the axionic mass $m_a$ i.e. $g_{ae}\approx 3\times (m_a/1~eV)$. Equation (\ref{eqn0}) can be remodeled as a spin system in a magnetic field e.g. $\mathbf{B_a}\cdot\mathbf{S}$, where $\mathbf{B_a}=\frac{g_{ae}}{f_a} \nabla a$. The effective axionic magnetic field \cite{BARBIERI2017135,Knirck_2018,SikivieRMP2021} along the $x~$axis is $B_{a_x}=B_{a_x}^0 \sin{\omega_a t}$, where $B_{a_x}^0=\frac{g_{ae}}{2q_e}\sqrt{\frac{n_a \hbar m_a}{c}} v_{x}$, $c$ is the speed of light in vacuum, $\hbar$ is the Planck constant, $q_e$ is the electronic charge, $n_a\approx 3\times 10^{8} (1~eV/m_a)$ is the axion dark matter density,  $v_{x}=v_a[b_0\cos{(\lambda_{lab})}-b_1\sin{(\lambda_{lab})}\cos{(\omega_d t+\phi_{lab}+\psi})]$ with $\omega_d=2\pi/T_d$ and $T_d\approx 84600~$s. The speed of axion $v_a$ in the laboratory frame varies between $220~$km/s and $250~$km/s in a period equal to a year \cite{Knirck_2018}. Parameters $b_0$, $b_1$ and $\psi$ vary over the year and depend on $v_a$ \cite{Knirck_2018}. The latitude and the longitude of the earth bound laboratory are denoted by $\lambda_{lab}$ and $\phi_{lab}$, respectively.

%%$n_a$ is the axion dark matter density, $g_{ae}$ is the axion-electron coupling strength, $q_e$ is the electronic charge, $h$ is the Planck constant, $c$ is the speed of light in vacuum, and $m_a$ is the mass of axion to be detected.  In the traditional Faraday configuration \cite{DillonJAP1963}, the propagation direction of the laser and the direction of the saturation magnetization are parallel.

Consider a ferromagnetic rod of length $L$ and radius $r$ is immersed in the axionic magnetic field $B_{a_x}$ and placed inside an optical cavity in a cryogenic environment. The cavity can be formed by using two identical high reflectivity mirrors as shown in Fig. \ref{fig1} or by appropriately coating all surfaces of the rod with anti-reflection coatings. Using the rod itself as a cavity can potentially render a simpler and more stable experimental setup. The rod is illuminated with a linearly polarized (along $z~$axis) probe light propagating along the cavity axis ($y~$axis, Fig. \ref{fig1}). The radius of the laser beam is $r$. The rod is magnetized to saturation using a dc magnetic field $B_0$ applied along the $z~$axis. An oscillating magnetic field, here the axionic field $B_{a_x}$, in the $x-y$ plane initiates a precession of the magnetic moment around $B_0$. The frequency of the precession $\omega_a\approx m_ac^2/\hbar$ and thus the mass of the dark matter to be detected is determined by the magnitude of $B_0$ via $\omega_a/2\pi= \gamma B_0$. For a rod, the ferromagnetic resonance frequency is slightly different than $\gamma B_0$ due to the shape related demagnetization \cite{KittelSolidState}. In the following, however, we do not consider this effect for the analytical simplicity. Precession of the magnetic moment around the $z-$axis means that a component of the magnetic moment appears in the $x-y~$plane. This component of the magnetic moment oscillates at $\omega_a$ and is known as the uniform precession mode or the Kittel mode.

The induced magnetic moment in the $x-y$ plane can be found by solving the Bloch equation \cite{KittelSolidState} $\frac{d\mathbf{m}}{dt}=\gamma( \mathbf{m}\times\mathbf{B})-\mathbf{m}/T_{1,2}$, where $\mathbf{m}$ is the magnetic moment of the rod, $\gamma$ is the gyro magnetic ratio, and $\mathbf{B}=B_{a_x}\hat{x}+B_0\hat{z}$. The spin-lattice and the spin-spin relaxation time are denoted by $T_1$ and $T_2$, respectively. Given $B_{a_x}$ is expected to be many orders of magnitude smaller than $B_0$ ($B_{a_x}\ll B_0$) \cite{AlesiniPRD2021}, we have $(m_x,~m_y)\ll m_z$. This allows us to set $m_z\approx m_s$ and $\frac{dm_z}{dt}\approx 0$, where $m_s$ is the saturation magnetic moment of the rod. In the steady state the $y$ component of the induced magnetic moment is

\begin{eqnarray}
m_y&=& \frac{m_s \gamma T_2}{2}B_{a_x}^0 \sin{\omega_at}. 
%m_y&=& \frac{m_s \gamma T_2}{2}B_{a_x} \sin{\omega_at}.
%m_y&=& m_s \gamma T_2(B_{a_x}\cos{\omega_at}+B_{a_y} \sin{\omega_at})/2.
%%\nonumber
%%m_y&=&\frac{\gamma m_s\sqrt{4\omega_a^2+\alpha^2 \omega_a^2}}{2\alpha\omega_a}B_{y}\sin{\omega_at}\\
%\nonumber
% m_y&=&\frac{ 2m_s \gamma B_{y}}{\alpha\sqrt{4+\alpha^2}\omega_a}\sin{\omega_at}\\
% \nonumber
% &\approx&\frac{m_s \gamma B_{y}}{\alpha\omega_a}\sin{\omega_at},\\
%%m_y&=&\frac{\mu_0M_s}{B_0^2-\omega_a^2/\gamma^2}\Bigl[B_0B_{a_y}-i \frac{\omega_a}{\gamma}B_{a_x}\Bigr]
%%m_y&=&\frac{\mu_0m_s}{B_0^2-\omega_a^2/\gamma^2} \sqrt{B_0^2B_{a_y}^2+\frac{\omega_a^2}{\gamma^2}B_{a_x}^2}
%%v_a&=&\sqrt{|(v_{LSR}+v_{pec})|^2+
\label{eqn2}
\end{eqnarray}

%% The Verdet constant is normally measured with $\mathbf{m_s}\parallel \mathbf{k}$ .

Linearly polarized light $\mathbf{E}_i=E_i\hat{z}$ propagating along the cavity axis ($y~$axis) interacts with $m_y$ via the Faraday effect and its plane of polarization rotates by an angle $\frac{V_rLm_y}{m_s}$  \cite{Deeter1990,HisatomiPRB2016,DillonJAP1963}, where $E_i=\sqrt{2I_i/(\epsilon_0c)}$, $\epsilon_0$ is the permittivity of free space, $I_i=\frac{P}{\pi r^2}$ is the intensity of the input laser light to the cavity, $P$ is the input optical power, and $V_r$ is the Verdet constant of the rod in radian per meter. The direction of the light field propagation is denoted by $\mathbf{k}=k\hat{y}$. To ensure that light of the purest polarization state enters the cavity, it is filtered using a polarizer of adequate extinction ratio. In principle, a polarizer of an arbitrarily high extinction ratio can be engineered by using multiple metamaterial polarizers in cascade \cite{Hemmati:19,denBoer:95}. In an optical isolator, also known as the Faraday rotator, the saturation magnetization $\mathbf{m_s}$ and the direction of the light field propagation are parallel and there is no magnetization orthogonal to $\mathbf{m_s}$ hence the rotation angle is $V_rL$ \cite{Deeter1990}. In our case, since only a fraction of the magnetization is parallel to $\mathbf{k}$, the rotation angle is reduced by the factor $\frac{m_y}{ms}$ \cite{Deeter1990,HisatomiPRB2016}. The presence of the optical cavity means that light bounces back and forth between the mirrors for $N=\tau_c/\tau$ times before exiting the cavity \cite{QuantumOpticsFox}, where $\tau_c=\frac{n_cL_{c}}{c(1-R_c)}$ is the photon lifetime in the cavity and $\tau=n_cL_{c}/c$ is the time required by the light to go from one mirror of the cavity to the other, $n_c$ is the refractive index of the cavity filler material, $L_c$ is the length of the cavity and $R_c$ is the reflectivity of the cavity mirrors. This means that the light interacts with $m_y$ $N$ times. Due to the non-reciprocal nature of the Faraday effect \cite{DillonJAP1963,Deeter1990}, each time light interacts with $m_y$, its plane of polarization is rotated by another $\frac{V_rLm_y}{m_s}$ radians. On exiting the cavity, the plane of polarization of the light field is rotated by an angle $\theta_f=\frac{NV_rm_y  L }{m_s}$. In an ideal lossless cavity, the electric field after the cavity is $\mathbf{E}_o=E_i\sin\theta_f \hat{x}+E_i\cos\theta_f\hat{z}$. The $x$ component ($E_x$) of $\mathbf{E}_o$ is picked up using a polarizer (Fig. \ref{fig1}) and is the desired signal. The optical power $P_x=\pi r^2 \epsilon_0 c \langle E_x^2\rangle$ is

\begin{eqnarray}
%\nonumber
%%P_x&=&\epsilon_0 c \langle E_x^2\rangle= I_0\sin^2\theta_f\\
P_x&=&P\sin^2\theta_f
\approx \frac{P V_r^2L^2\gamma^2T_2^2}{4(1-R_c)^2}(B_{a_x}^0\sin{\omega_at})^2,
\label{eqn3}
\end{eqnarray}

where we have set $\sin\theta_f\approx \theta_f$ as $\theta_f$ is expected to be $\ll1^\circ$ (see below). Equation (\ref{eqn3}) shows that $P_x$ is subjected to the daily and the yearly modulation via the axionic field $B_{a_x}$, a unique signature of the axion dark matter. Most importantly, however, axions of any mass, at least in principle, can be searched by merely adjusting the dc magnetic field ($\omega_a/2\pi=\gamma B_0$) without changing any component of the proposed experiment. This is possible as axions of different masses only change the plane of polarization of the light field. Consequently, the optical cavity remains on resonance with the laser when $B_0$ is changed. Since $B_0$ can be controlled by changing the current through the magnet, no component in the setup needs to be modified for searching different axions. Hence the whole axion spectrum within the sensitivity of the Faraday haloscope can be searched using the same setup in a relatively short time (see below). The axion induced signal $P_x$ can be enhanced using a long rod, a higher optical power, a long spin coherence time and a magnetic material with a high Verdet constant. $P_x$ can also be boosted by increasing $N=1/(1-R_c)$ using ultra high reflectivity cavity mirrors provided $N\tau\le T_2$. This implies that the preferable cavity length is $L_c=L$.

%%% multi layer polarizer https://www.jpier.org/issues/volume.html?paper=21031107

%% Note that all information about the magnetic material e.g. the spin density is contained in the Verdet constant. The spectral resolution may be limited by the magnet ramp up and settling down time and the time required to acquire sufficient data for averaging.  For example, by selecting any eight axion mass within the theoretical axion spectrum, data can be collected for three hours for the each setting and should provide sufficient averaging time. This is another significant advantage of the current scheme over others which require years \cite{AdmxPRL2021,AlesiniPRD2021,SikivieRMP2021}. This is in stark contrast with other microwave cavity based axion search schemes which require modifications to the cavity lengths/parameters for searching axion of different mass \cite{FLOWER2019100306,AlesiniPRD2021,AdmxPRL2021}.  equivalent to a power $h\omega$ W, where $\omega$ 

Detection is often the most challenging aspect of axion search experiments. Here we accomplish this using a photon counter which is capable of detecting a single photon. High quantum efficiency ($\approx 1$) photon counters in the visible \cite{Ceccarelli2021} and the infrared bands \cite{ZwillerAPL2021,SchweickertAPL2018} are readily available. Indeed, it is a unique feature of the proposed experiment in that it converts axions into optical photons and then takes advantage of the high efficiency photon counters available in the optical domain. Photon counters in the infrared band have dark counts, a potential source of noise, $\ll 1~s^{-1}$ \cite{ZwillerAPL2021,SchweickertAPL2018}. As a result, detection can be accomplished free of noise which often, in existing experiments, dominates axionic signals \cite{AdmxPRL2021,CresciniPRL2020,FLOWER2019100306}. The number of photons equivalent to $P_x$ is $P_x\lambda/(hc)$, where $\lambda$ is the wavelength of the laser light.
  
  %% Note that photon counters do not require additional amplification.

 %% This is a significant advantage of the proposed experiment over other experiments \cite{AlesiniPRD2021,AdmxPRL2021} which detect microwaves for searching axion but cannot count microwave photons easily since such counters are not readily available.
 
 %%The ability of counting photons in the optical domain  This detection scheme along with the ability of searching a wide range of axion mass using the same setup are the salient features of the current proposal.

%%With a photon counter of quantum efficiency $\eta$, the number of photons detected is $N_{ph}= \frac{\eta\lambda P_x}{h c}$. 

\begin{figure}
    \centering
    \includegraphics[width=8.6cm]{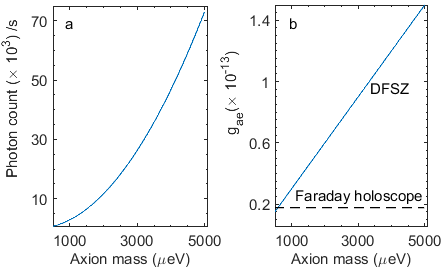}
    \caption{a) Assuming the DFSZ model of axion-electron coupling, the average number of photons detected as a function of axion mass at midnight on January $1$ when a $r=12.5~$mm and $L=100~$mm cerium doped YIG rod is illuminated with $100~$mW laser power. Other parameters are $R_c=99.999\%$, $\lambda=1550~$nm, $T_2=50~\mu s$ and $V_r=1.75\times 10^4~$rad./m. b) Axion electron coupling strength $g_{ae}$ as a function of axion mass in DFSZ model (solid line). The dashed line shows the minimum $g_{ae}$ that the Faraday haloscope can operate with and produces at least $1000$ photon counts (see the main text for detail) . }
    \label{fig2}
\end{figure}

%% With a beam waist radius $12.5~$mm, the intensity of the laser is $2\times 10^2~$W m$^{-2}$ outside the cavity and $2\times 10^7~$W m$^{-2}$ inside the cavity.
%The magnetic moment of YIG arises from the spins of the four uncompensated Fe$^{3+}$ ions found per primitive cell in a YIG lattice \cite{Serga_2010}.

Let us now consider an example. Here, a magnetic material with a high Verdet constant and a long spin coherence time is desirable. Yttrium iron garnet (YIG) \cite{HisatomiPRB2016} and its variants such as cerium substituted YIG \cite{LageAPL2017,HIGUCHI2003293} might be good candidates. YIG is widely used in quantum magnonics \cite{Lachance_Quirion_2019} and science \cite{Rahman_2019} for its long spin coherence time. It has a moderate Verdet constant \cite{HisatomiPRB2016} and is often used as an optical isolator meaning that it can handle high intensity lasers \cite{Higuchi_1999,LageAPL2017}. Substitution of yttrium by the non-magnetic cerium (Ce) enhances the Faraday rotation of YIG significantly while maintaining the magnetic properties of the undoped YIG \cite{HIGUCHI2003293,Higuchi_1999}. At cryogenic temperature, $V_r$ of Ce:YIG increases rapidly and at $300~$mK in a cryostat it can reach $\approx 1.75\times 10^4~$rad/m \cite{LageAPL2017}. Figure \ref{fig2} shows the average number of photons as a function of axion mass when a $r=12.5~$mm and $L=100~$mm Ce:YIG rod is illuminated with $100~$mW optical power at $\lambda=1550~$nm. At this wavelength cerium doped YIG is transparent \cite{Higuchi_1999}. Lasers with extremely low intensity and frequency noise, narrow linewidths ($\approx 1~$kHz) and high stability at $1550~$nm are available \cite{RioGrande1550nm,MeylahnPRD2022}. The $\langle 111\rangle$ crystalline axis of YIG is along the long axis of the rod. The strength of the required magnetic field for scanning the axion mass range shown in Fig. \ref{fig2}a is between $B_0=4.3~$T and $42.85~$T. Highly homogeneous superconducting magnets up to $28.2~$T are commercially available \cite{CallonMorgane2021} and a $45~$T magnet is available as a national facility for decades \cite{HahnSeungyong2019,MillerIEEE2003}. At $m_a=500~\mu$eV, the number of photons is $\approx 1\times 10^3~$/s while at $2000~\mu$eV it is $\approx 12 \times 10^3~$/s. These are significant and should be easily detectable using a $1550~$nm photon counter \cite{ZwillerAPL2021,SchweickertAPL2018}. Using a higher laser power, axions below $500~\mu$eV can be searched. At $500~\mu$eV, $\theta_f$ is $2.20\times 10^{-8}~$rad. In the calculations we have used the co-ordinates of the University of Warwick, $R_c=99.999\%$ \cite{RempeOL1992} and a spin coherence time of $T_2=50~\mu$s \cite{Spencer1961,KittelSolidState}. Note that $T_2=1/\Delta f$, where $\Delta f$ is the full width half maximum linewidth of a magnetic resonance peak \cite{KittelSolidState}. 

%% In ferromagnetic systems, traditionally, linewidths are measured which are inversely proportional to $T_2$ \cite{KittelSolidState}.  and parallel to the $y~$axis (Fig. \ref{fig1})  
%%A non-ideal polarization extinction of polarizers particular that may degrade the performance of the Faraday haloscope. 
%%Sources of noise that may degrade the overall photon count include intensity fluctuation of the laser, counts due to the finite extinction ratios ($\gamma_r$) of the polarizers and the detector dark counts. For the noise free operation of the Faraday haloscope  operated in free ru   Fluctuation in the laser power due to shot noise,  Another potential source of The proposed experiment requires a prolonged period of operation and a long averaging time. As a result, a highly stable laser is essential. Fortunately, such a laser system has been recently developed \cite{MeylahnPRD2022}. 

%%Sources of noise that may degrade the performance of the Faraday haloscope intensity fluctuation of the laser, counts due to the finite extinction ratios ($\gamma_r$) of the polarizers and the detector dark counts.

To ensure that light polarized along the $z~$axis (Fig. \ref{fig1}) does not reach the detector, the polarizer in front of the photon counter must efficiently block the $z~$ polarized light. In principle, a polarizer can have an arbitrarily high polarization extinction ratio $\gamma_r$ \cite{Hemmati:19,denBoer:95}. The number of photons in the $1550~$nm beam ($100~$mW) is $\approx 10^{18}$. To ensure that none of these photons enters the photon counter when no axion ($E_x=0$) is present, $\gamma_r=10^{20}$ is sufficient. To achieve this level of polarization extinction up to eight dual grating metamaterial based polarizers may be required \cite{Hemmati:19}. Importantly, light transmission through these polarizers is close to $100\%$. The relative intensity noise (RIN) which represents the combined effect of shot noise, $1/f$ noise and beam pointing instabilities can potentially cause performance degradation of the Faraday haloscope via fluctuation in the photon counts. However, highly stable, low intensity and frequency noise, and narrow linewidths lasers are available \cite{RioGrande1550nm,MeylahnPRD2022}. When monitored for a prolonged period of time ($\approx 2~$weeks) such a laser showed virtually no drift \cite{MeylahnPRD2022}. This laser has a RIN $\le -140~$dB/Hz at frequencies $\ge 0~$ Hz. When integrated between dc and $1~$MHz assuming a constant RIN of $-140~$dB/Hz, such a noise results in an intensity fluctuation and hence a fluctuation in the photon count of $<0.1\%$. This is negligible. Furthermore, the effect of laser shot noise on the photon count can be reduced by using the amplitude squeezed light \cite{AasiJ2013Esot}. Performance degradation of the Faraday haloscope due to the dark count of the photon counter is also negligible as such a counter has a very low dark count e.g. $5.6\times 10^{-3}$ per second \cite{SchweickertAPL2018}. Unwanted photons produced by the thermal excitation of the Kittel magnon can also reduce the fidelity of the axion produced photon counts. Here, a high purity sample is essential so that a very little or no laser power is absorbed. Any residual heating due to laser and blackbody absorption by the rod should be actively mitigated by the cryogenics with a sufficient cooling power. Nevertheless, the magnetic moment along the $y$ axis due to such potential stimuli at temperature $T$ \cite{KittelSolidState} is $m_y^{th}=\mu_B V D(\omega)/(e^{\hbar\omega_a/k_BT}-1)$, where $\mu_B$ is the Bohr magneton, $D(\omega_a)=(\frac{\hbar}{2JSa^2})^{3/2}\sqrt{\omega_a/16\pi^2}$ is the density of states of magnon at $\omega_a$, $J$ is the ferromagnetic exchange integral, $a$ is the lattice constant and $V$ is the volume of the rod. At $T=300~$mK and $m_a=500~\mu$eV ($\omega_a/2\pi=120~$GHz), we have $m_y^{th}=1.63\times 10^{-22}~$A/m. This is trivial compared to $\langle m_y\rangle=5\times 10^{-16}~$A/m (Eq. (\ref{eqn2})). Consequently, photon counts originating from thermal sources are not relevant. 

Taking at least $1\times 10^3$ photons per second ($\gg$ the dark count) due to axion is a sufficiently strong signal, Fig. \ref{fig2}b shows the minimum axion-electron coupling strength that the Faraday haloscope can operate with. For comparison, this figure also shows $g_{ae}$ expected from the DFSZ model. It is clear that the Faraday haloscope can produce enough photon counts even when the axion-electron coupling strength is smaller than that is predicted by the DFSZ model.

\begin{figure}
    \centering
    \includegraphics[width=8.5cm]{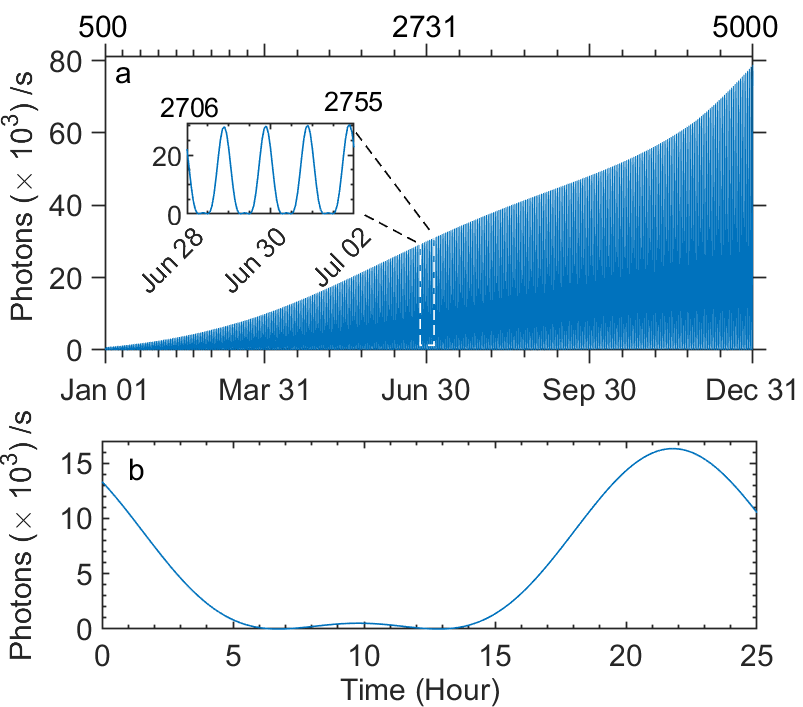}
    \caption{a) Photon counts when axions of different mass are searched for over a year. The top scale shows the axionic mass. The inset is the zoomed view when the search reaches $2731~\mu$eV (Jun $30$). b) The predicted daily modulation of the number of photons counted over a day when the search is fixed at $m_a=2000~\mu$eV (May $02$).}
    \label{fig3}
\end{figure}

The rate at which axions can be searched for is critically important. In the proposed experiment, the magnetic field is the only parameter that needs to be changed for searching different axions and can be accomplished by changing the current through the magnet. Assuming twenty data points per day, each $0.6164~\mu$eV apart, axions between $500$ and $5000~\mu$eV can be searched for in a year. Figure \ref{fig3}a shows the result of such a search starting on January $1$ ($500~\mu$eV) and ending on December $31$ ($5000~\mu$eV). Due to the daily and yearly motion of the earth through the dark matter halo, the photon count oscillates (inset, Fig. \ref{fig3}a) as the search progresses. In particular, due to the earth's rotational motion, irrespective of the axion mass, the signal goes to zero on daily basis at certain time of the day. This is captured in more detail in Fig. \ref{fig3}b where the search is fixed at $2000~\mu$eV (May $02$). It is clear that at around $10~$am the photon count is minimal while at around midnight the expected photon count peaks. Importantly, these periodic modulations mean that the proposed experiment, at least in principle, should be immune to any background count that may arise. In particular, due to the rotational and yearly motion of the earth the photon count arising from axions oscillates periodically (Fig. \ref{fig3}). In contrast, counts due to the background, on the average, should remain fixed and as such should appear as an offset in the measurement process.

As a complementary technique to photon counting, the anti-Stokes sideband associated with the Brillouin light scattering can be considered \cite{HisatomiPRL2019}. In this configuration, circularly polarized light is used and the optical cavity is avoided. The external magnetic field $B_0$ is applied along the $y$ axis (Fig. \ref{fig1}). The axionic microwave field excites magnetization/magnons in the $x-z~$plane. The interaction \cite{HisatomiPRL2019} between a Kittel magnon (zero linear momentum) and the circularly polarized light creates an anti-Stokes sideband at $\omega_l+\omega_a$, where $\omega_l=2\pi \lambda/c$ is the angular frequency of the laser. The conservation of the angular momentum in the process of magnon (axion) to photon conversion is guaranteed via the rotational recoil in the crystal \cite{HisatomiPRL2019}. Also, since the light travelling through the Faraday active material interacts with the magnon only once, even a small $T_2$ can be sufficient as long as $T_2>Ln_c/c$. This can be useful in using other materials with higher Verdet constants such as chromium tribromide \cite{DillonJAP1963}. Importantly, this detection scheme can provide spectroscopic information e.g. the frequency of the axion detected and its linewidth. The as-produced sideband interferes with a strong local oscillator at frequency $\omega_{lo}$ in a heterodyne scheme on a fast photodiode \cite{HisatomiPRL2019}. We have $[E_{lo}\cos{\omega_{lo}t}+E_s\cos{(\omega_l+\omega_a)t}]^2$, where $E_s$ is the amplitude of the scattered light and $E_{lo}\cos\omega_{lo}t$ is the local oscillator field. The signal of interest is now $E_{lo}E_s\cos{(\omega_l-\omega_{lo}+\omega_a)t}$, where we have not shown the dc offset and the high frequency terms. As before, the frequency of axions to be detected is selected by adjusting the dc magnetic field (Fig. \ref{fig1}). A significant advantage of this interferometric detection is that a weak signal can be strengthened with a strong local oscillator. Given the availability of a low RIN (low amplitude noise), highly stable, narrow linewidth ($\approx 1~$kHz) and low frequency/phase noise laser (e.g. $0.7~\mu rad./\sqrt{Hz}$ at $1~$kHz) \cite{RioGrande1550nm,MeylahnPRD2022}, the heterodyne scheme proposed here is robust. 

Finally, it is instructive to consider the sensitivity range of the Faraday haloscope and its place in the global axion search landscape. The cavity based search experiments such as the ADMX \cite{AdmxPRL2021} and the QUAX \cite{AlesiniPRD2021,CresciniPRL2020}  collaborations can search axions up to few tens of $\mu$eV while the proposed MADMAX experiment \cite{MadmaxPRL2017} is sensitive between the axion mass $40$ and $400~\mu$eV. The proposed nuclear resonance \cite{AravanitakiPRL2014} and the topological insulator \cite{MarshPRL2019} based schemes are sensitive to $1$ to $1000~\mu$eV and $700$ to $3500~\mu$eV, respectively. The here proposed Faraday haloscope can search axions between $500$ and $5000~\mu$eV and is uniquely placed to expand the global axion search above $3500~\mu$eV.

%% is focused in the mass range $1$ to $10~\mu$eV while the latest experimental results of the QUAX collaboration  is centered around $43~\mu$eV.

In conclusion, we have theoretically shown that axions of mass between $500~\mu$eV and $5000~\mu$eV can be searched for using a single experimental setup. In realizing the search scheme we have used the axion-induced rotation of the plane of polarization of the probe light field via the Faraday effect. Using an optical cavity has allowed us to amplify the weak Faraday effect significantly. We have also shown that a wideband search of axions over hundreds of $\mu$eV can be accomplished in the matter of a year. It has also been demonstrated that the detection can be accomplished using high quantum efficiency, well developed and readily available photon counters in almost a noise free environment. In the current article, we have considered cerium doped YIG as the Faraday active material which has a moderate Verdet constant. In contrast, chromium tribromide \cite{DillonJAP1963} has a significantly higher Verdet constant and can potentially be used instead of YIG. This can enhance the optical signal in a major way.

%%Acknowledgment: I acknowledge the comments and suggestions that I have received from G. Morley, C. Ghag and Y. Ramachars which have improved the manuscript.

%%\bibliography{main.bib}

\begin{thebibliography}{44}%
\makeatletter
\providecommand \@ifxundefined [1]{%
 \@ifx{#1\undefined}
}%
\providecommand \@ifnum [1]{%
 \ifnum #1\expandafter \@firstoftwo
 \else \expandafter \@secondoftwo
 \fi
}%
\providecommand \@ifx [1]{%
 \ifx #1\expandafter \@firstoftwo
 \else \expandafter \@secondoftwo
 \fi
}%
\providecommand \natexlab [1]{#1}%
\providecommand \enquote  [1]{``#1''}%
\providecommand \bibnamefont  [1]{#1}%
\providecommand \bibfnamefont [1]{#1}%
\providecommand \citenamefont [1]{#1}%
\providecommand \href@noop [0]{\@secondoftwo}%
\providecommand \href [0]{\begingroup \@sanitize@url \@href}%
\providecommand \@href[1]{\@@startlink{#1}\@@href}%
\providecommand \@@href[1]{\endgroup#1\@@endlink}%
\providecommand \@sanitize@url [0]{\catcode `\\12\catcode `\$12\catcode
  `\&12\catcode `\#12\catcode `\^12\catcode `\_12\catcode `\%12\relax}%
\providecommand \@@startlink[1]{}%
\providecommand \@@endlink[0]{}%
\providecommand \url  [0]{\begingroup\@sanitize@url \@url }%
\providecommand \@url [1]{\endgroup\@href {#1}{\urlprefix }}%
\providecommand \urlprefix  [0]{URL }%
\providecommand \Eprint [0]{\href }%
\providecommand \doibase [0]{https://doi.org/}%
\providecommand \selectlanguage [0]{\@gobble}%
\providecommand \bibinfo  [0]{\@secondoftwo}%
\providecommand \bibfield  [0]{\@secondoftwo}%
\providecommand \translation [1]{[#1]}%
\providecommand \BibitemOpen [0]{}%
\providecommand \bibitemStop [0]{}%
\providecommand \bibitemNoStop [0]{.\EOS\space}%
\providecommand \EOS [0]{\spacefactor3000\relax}%
\providecommand \BibitemShut  [1]{\csname bibitem#1\endcsname}%
\let\auto@bib@innerbib\@empty
%</preamble>
\bibitem [{\citenamefont {Sikivie}(2021)}]{SikivieRMP2021}%
  \BibitemOpen
  \bibfield  {author} {\bibinfo {author} {\bibfnamefont {P.}~\bibnamefont
  {Sikivie}},\ }\bibfield  {title} {\bibinfo {title} {Invisible axion search
  methods},\ }\href {https://doi.org/10.1103/RevModPhys.93.015004} {\bibfield
  {journal} {\bibinfo  {journal} {Rev. Mod. Phys.}\ }\textbf {\bibinfo {volume}
  {93}},\ \bibinfo {pages} {015004} (\bibinfo {year} {2021})}\BibitemShut
  {NoStop}%
\bibitem [{\citenamefont {Peccei}\ and\ \citenamefont
  {Quinn}(1977)}]{PecceiQuinn1977}%
  \BibitemOpen
  \bibfield  {author} {\bibinfo {author} {\bibfnamefont {R.~D.}\ \bibnamefont
  {Peccei}}\ and\ \bibinfo {author} {\bibfnamefont {H.~R.}\ \bibnamefont
  {Quinn}},\ }\bibfield  {title} {\bibinfo {title} {$\mathrm{CP}$ conservation
  in the presence of pseudoparticles},\ }\href
  {https://doi.org/10.1103/PhysRevLett.38.1440} {\bibfield  {journal} {\bibinfo
   {journal} {Phys. Rev. Lett.}\ }\textbf {\bibinfo {volume} {38}},\ \bibinfo
  {pages} {1440} (\bibinfo {year} {1977})}\BibitemShut {NoStop}%
\bibitem [{\citenamefont {Preskill}\ \emph {et~al.}(1983)\citenamefont
  {Preskill}, \citenamefont {Wise},\ and\ \citenamefont
  {Wilczek}}]{PRESKILL1983127}%
  \BibitemOpen
  \bibfield  {author} {\bibinfo {author} {\bibfnamefont {J.}~\bibnamefont
  {Preskill}}, \bibinfo {author} {\bibfnamefont {M.~B.}\ \bibnamefont {Wise}},\
  and\ \bibinfo {author} {\bibfnamefont {F.}~\bibnamefont {Wilczek}},\
  }\bibfield  {title} {\bibinfo {title} {Cosmology of the invisible axion},\
  }\href {https://doi.org/https://doi.org/10.1016/0370-2693(83)90637-8}
  {\bibfield  {journal} {\bibinfo  {journal} {Phys. Lett. B}\ }\textbf
  {\bibinfo {volume} {120}},\ \bibinfo {pages} {127} (\bibinfo {year}
  {1983})}\BibitemShut {NoStop}%
\bibitem [{\citenamefont {Dine}\ and\ \citenamefont
  {Fischler}(1983)}]{DINE1983137}%
  \BibitemOpen
  \bibfield  {author} {\bibinfo {author} {\bibfnamefont {M.}~\bibnamefont
  {Dine}}\ and\ \bibinfo {author} {\bibfnamefont {W.}~\bibnamefont
  {Fischler}},\ }\bibfield  {title} {\bibinfo {title} {The not-so-harmless
  axion},\ }\href
  {https://doi.org/https://doi.org/10.1016/0370-2693(83)90639-1} {\bibfield
  {journal} {\bibinfo  {journal} {Phys. Lett. B}\ }\textbf {\bibinfo {volume}
  {120}},\ \bibinfo {pages} {137} (\bibinfo {year} {1983})}\BibitemShut
  {NoStop}%
\bibitem [{\citenamefont {Abbott}\ and\ \citenamefont
  {Sikivie}(1983)}]{ABBOTT1983133}%
  \BibitemOpen
  \bibfield  {author} {\bibinfo {author} {\bibfnamefont {L.}~\bibnamefont
  {Abbott}}\ and\ \bibinfo {author} {\bibfnamefont {P.}~\bibnamefont
  {Sikivie}},\ }\bibfield  {title} {\bibinfo {title} {A cosmological bound on
  the invisible axion},\ }\href
  {https://doi.org/https://doi.org/10.1016/0370-2693(83)90638-X} {\bibfield
  {journal} {\bibinfo  {journal} {Phys. Lett. B}\ }\textbf {\bibinfo {volume}
  {120}},\ \bibinfo {pages} {133} (\bibinfo {year} {1983})}\BibitemShut
  {NoStop}%
\bibitem [{\citenamefont {Barbieri}\ \emph {et~al.}(2017)\citenamefont
  {Barbieri}, \citenamefont {Braggio}, \citenamefont {Carugno}, \citenamefont
  {Gallo}, \citenamefont {Lombardi}, \citenamefont {Ortolan}, \citenamefont
  {Pengo}, \citenamefont {Ruoso},\ and\ \citenamefont
  {Speake}}]{BARBIERI2017135}%
  \BibitemOpen
  \bibfield  {author} {\bibinfo {author} {\bibfnamefont {R.}~\bibnamefont
  {Barbieri}}, \bibinfo {author} {\bibfnamefont {C.}~\bibnamefont {Braggio}},
  \bibinfo {author} {\bibfnamefont {G.}~\bibnamefont {Carugno}}, \bibinfo
  {author} {\bibfnamefont {C.}~\bibnamefont {Gallo}}, \bibinfo {author}
  {\bibfnamefont {A.}~\bibnamefont {Lombardi}}, \bibinfo {author}
  {\bibfnamefont {A.}~\bibnamefont {Ortolan}}, \bibinfo {author} {\bibfnamefont
  {R.}~\bibnamefont {Pengo}}, \bibinfo {author} {\bibfnamefont
  {G.}~\bibnamefont {Ruoso}},\ and\ \bibinfo {author} {\bibfnamefont
  {C.}~\bibnamefont {Speake}},\ }\bibfield  {title} {\bibinfo {title}
  {Searching for galactic axions through magnetized media: The quax proposal},\
  }\href {https://doi.org/https://doi.org/10.1016/j.dark.2017.01.003}
  {\bibfield  {journal} {\bibinfo  {journal} {Phys. Dark Universe}\ }\textbf
  {\bibinfo {volume} {15}},\ \bibinfo {pages} {135} (\bibinfo {year}
  {2017})}\BibitemShut {NoStop}%
\bibitem [{\citenamefont {Bartram~et al.}(2021)}]{AdmxPRL2021}%
  \BibitemOpen
  \bibfield  {author} {\bibinfo {author} {\bibfnamefont {C.}~\bibnamefont
  {Bartram~et al.}} (\bibinfo {collaboration} {ADMX Collaboration}),\
  }\bibfield  {title} {\bibinfo {title} {Search for invisible axion dark matter
  in the $3.3--4.2\text{ }\text{ }\ensuremath{\mu}\mathrm{eV}$ mass range},\
  }\href {https://doi.org/10.1103/PhysRevLett.127.261803} {\bibfield  {journal}
  {\bibinfo  {journal} {Phys. Rev. Lett.}\ }\textbf {\bibinfo {volume} {127}},\
  \bibinfo {pages} {261803} (\bibinfo {year} {2021})}\BibitemShut {NoStop}%
\bibitem [{\citenamefont {Du}\ \emph {et~al.}(2018)\citenamefont {Du},
  \citenamefont {Force}, \citenamefont {Khatiwada}, \citenamefont {Lentz},
  \citenamefont {Ottens}, \citenamefont {Rosenberg}, \citenamefont {Rybka},
  \citenamefont {Carosi}, \citenamefont {Woollett}, \citenamefont {Bowring},
  \citenamefont {Chou}, \citenamefont {Sonnenschein}, \citenamefont {Wester},
  \citenamefont {Boutan}, \citenamefont {Oblath}, \citenamefont {Bradley},
  \citenamefont {Daw}, \citenamefont {Dixit}, \citenamefont {Clarke},
  \citenamefont {O'Kelley}, \citenamefont {Crisosto}, \citenamefont {Gleason},
  \citenamefont {Jois}, \citenamefont {Sikivie}, \citenamefont {Stern},
  \citenamefont {Sullivan}, \citenamefont {Tanner},\ and\ \citenamefont
  {Hilton}}]{DuPRL2018}%
  \BibitemOpen
  \bibfield  {author} {\bibinfo {author} {\bibfnamefont {N.}~\bibnamefont
  {Du}}, \bibinfo {author} {\bibfnamefont {N.}~\bibnamefont {Force}}, \bibinfo
  {author} {\bibfnamefont {R.}~\bibnamefont {Khatiwada}}, \bibinfo {author}
  {\bibfnamefont {E.}~\bibnamefont {Lentz}}, \bibinfo {author} {\bibfnamefont
  {R.}~\bibnamefont {Ottens}}, \bibinfo {author} {\bibfnamefont {L.~J.}\
  \bibnamefont {Rosenberg}}, \bibinfo {author} {\bibfnamefont {G.}~\bibnamefont
  {Rybka}}, \bibinfo {author} {\bibfnamefont {G.}~\bibnamefont {Carosi}},
  \bibinfo {author} {\bibfnamefont {N.}~\bibnamefont {Woollett}}, \bibinfo
  {author} {\bibfnamefont {D.}~\bibnamefont {Bowring}}, \bibinfo {author}
  {\bibfnamefont {A.~S.}\ \bibnamefont {Chou}}, \bibinfo {author}
  {\bibfnamefont {A.}~\bibnamefont {Sonnenschein}}, \bibinfo {author}
  {\bibfnamefont {W.}~\bibnamefont {Wester}}, \bibinfo {author} {\bibfnamefont
  {C.}~\bibnamefont {Boutan}}, \bibinfo {author} {\bibfnamefont {N.~S.}\
  \bibnamefont {Oblath}}, \bibinfo {author} {\bibfnamefont {R.}~\bibnamefont
  {Bradley}}, \bibinfo {author} {\bibfnamefont {E.~J.}\ \bibnamefont {Daw}},
  \bibinfo {author} {\bibfnamefont {A.~V.}\ \bibnamefont {Dixit}}, \bibinfo
  {author} {\bibfnamefont {J.}~\bibnamefont {Clarke}}, \bibinfo {author}
  {\bibfnamefont {S.~R.}\ \bibnamefont {O'Kelley}}, \bibinfo {author}
  {\bibfnamefont {N.}~\bibnamefont {Crisosto}}, \bibinfo {author}
  {\bibfnamefont {J.~R.}\ \bibnamefont {Gleason}}, \bibinfo {author}
  {\bibfnamefont {S.}~\bibnamefont {Jois}}, \bibinfo {author} {\bibfnamefont
  {P.}~\bibnamefont {Sikivie}}, \bibinfo {author} {\bibfnamefont
  {I.}~\bibnamefont {Stern}}, \bibinfo {author} {\bibfnamefont {N.~S.}\
  \bibnamefont {Sullivan}}, \bibinfo {author} {\bibfnamefont {D.~B.}\
  \bibnamefont {Tanner}},\ and\ \bibinfo {author} {\bibfnamefont {G.~C.}\
  \bibnamefont {Hilton}} (\bibinfo {collaboration} {ADMX Collaboration}),\
  }\bibfield  {title} {\bibinfo {title} {Search for invisible axion dark matter
  with the axion dark matter experiment},\ }\href
  {https://doi.org/10.1103/PhysRevLett.120.151301} {\bibfield  {journal}
  {\bibinfo  {journal} {Phys. Rev. Lett.}\ }\textbf {\bibinfo {volume} {120}},\
  \bibinfo {pages} {151301} (\bibinfo {year} {2018})}\BibitemShut {NoStop}%
\bibitem [{\citenamefont {Asztalos}\ \emph {et~al.}(2010)\citenamefont
  {Asztalos}, \citenamefont {Carosi}, \citenamefont {Hagmann}, \citenamefont
  {Kinion}, \citenamefont {van Bibber}, \citenamefont {Hotz}, \citenamefont
  {Rosenberg}, \citenamefont {Rybka}, \citenamefont {Hoskins}, \citenamefont
  {Hwang}, \citenamefont {Sikivie}, \citenamefont {Tanner}, \citenamefont
  {Bradley},\ and\ \citenamefont {Clarke}}]{AsztalozPRL2010}%
  \BibitemOpen
  \bibfield  {author} {\bibinfo {author} {\bibfnamefont {S.~J.}\ \bibnamefont
  {Asztalos}}, \bibinfo {author} {\bibfnamefont {G.}~\bibnamefont {Carosi}},
  \bibinfo {author} {\bibfnamefont {C.}~\bibnamefont {Hagmann}}, \bibinfo
  {author} {\bibfnamefont {D.}~\bibnamefont {Kinion}}, \bibinfo {author}
  {\bibfnamefont {K.}~\bibnamefont {van Bibber}}, \bibinfo {author}
  {\bibfnamefont {M.}~\bibnamefont {Hotz}}, \bibinfo {author} {\bibfnamefont
  {L.~J.}\ \bibnamefont {Rosenberg}}, \bibinfo {author} {\bibfnamefont
  {G.}~\bibnamefont {Rybka}}, \bibinfo {author} {\bibfnamefont
  {J.}~\bibnamefont {Hoskins}}, \bibinfo {author} {\bibfnamefont
  {J.}~\bibnamefont {Hwang}}, \bibinfo {author} {\bibfnamefont
  {P.}~\bibnamefont {Sikivie}}, \bibinfo {author} {\bibfnamefont {D.~B.}\
  \bibnamefont {Tanner}}, \bibinfo {author} {\bibfnamefont {R.}~\bibnamefont
  {Bradley}},\ and\ \bibinfo {author} {\bibfnamefont {J.}~\bibnamefont
  {Clarke}},\ }\bibfield  {title} {\bibinfo {title} {Squid-based microwave
  cavity search for dark-matter axions},\ }\href
  {https://doi.org/10.1103/PhysRevLett.104.041301} {\bibfield  {journal}
  {\bibinfo  {journal} {Phys. Rev. Lett.}\ }\textbf {\bibinfo {volume} {104}},\
  \bibinfo {pages} {041301} (\bibinfo {year} {2010})}\BibitemShut {NoStop}%
\bibitem [{\citenamefont {Alesini}\ \emph {et~al.}(2021)\citenamefont
  {Alesini}, \citenamefont {Braggio}, \citenamefont {Carugno}, \citenamefont
  {Crescini}, \citenamefont {D'Agostino}, \citenamefont {Di~Gioacchino},
  \citenamefont {Di~Vora}, \citenamefont {Falferi}, \citenamefont
  {Gambardella}, \citenamefont {Gatti}, \citenamefont {Iannone}, \citenamefont
  {Ligi}, \citenamefont {Lombardi}, \citenamefont {Maccarrone}, \citenamefont
  {Ortolan}, \citenamefont {Pengo}, \citenamefont {Rettaroli}, \citenamefont
  {Ruoso}, \citenamefont {Taffarello},\ and\ \citenamefont
  {Tocci}}]{AlesiniPRD2021}%
  \BibitemOpen
  \bibfield  {author} {\bibinfo {author} {\bibfnamefont {D.}~\bibnamefont
  {Alesini}}, \bibinfo {author} {\bibfnamefont {C.}~\bibnamefont {Braggio}},
  \bibinfo {author} {\bibfnamefont {G.}~\bibnamefont {Carugno}}, \bibinfo
  {author} {\bibfnamefont {N.}~\bibnamefont {Crescini}}, \bibinfo {author}
  {\bibfnamefont {D.}~\bibnamefont {D'Agostino}}, \bibinfo {author}
  {\bibfnamefont {D.}~\bibnamefont {Di~Gioacchino}}, \bibinfo {author}
  {\bibfnamefont {R.}~\bibnamefont {Di~Vora}}, \bibinfo {author} {\bibfnamefont
  {P.}~\bibnamefont {Falferi}}, \bibinfo {author} {\bibfnamefont
  {U.}~\bibnamefont {Gambardella}}, \bibinfo {author} {\bibfnamefont
  {C.}~\bibnamefont {Gatti}}, \bibinfo {author} {\bibfnamefont
  {G.}~\bibnamefont {Iannone}}, \bibinfo {author} {\bibfnamefont
  {C.}~\bibnamefont {Ligi}}, \bibinfo {author} {\bibfnamefont {A.}~\bibnamefont
  {Lombardi}}, \bibinfo {author} {\bibfnamefont {G.}~\bibnamefont
  {Maccarrone}}, \bibinfo {author} {\bibfnamefont {A.}~\bibnamefont {Ortolan}},
  \bibinfo {author} {\bibfnamefont {R.}~\bibnamefont {Pengo}}, \bibinfo
  {author} {\bibfnamefont {A.}~\bibnamefont {Rettaroli}}, \bibinfo {author}
  {\bibfnamefont {G.}~\bibnamefont {Ruoso}}, \bibinfo {author} {\bibfnamefont
  {L.}~\bibnamefont {Taffarello}},\ and\ \bibinfo {author} {\bibfnamefont
  {S.}~\bibnamefont {Tocci}},\ }\bibfield  {title} {\bibinfo {title} {Search
  for invisible axion dark matter of mass ${\mathrm{m}}_{a}=43\text{ }\text{
  }\ensuremath{\mu}\mathrm{eV}$ with the quax--$a\ensuremath{\gamma}$
  experiment},\ }\href {https://doi.org/10.1103/PhysRevD.103.102004} {\bibfield
   {journal} {\bibinfo  {journal} {Phys. Rev. D}\ }\textbf {\bibinfo {volume}
  {103}},\ \bibinfo {pages} {102004} (\bibinfo {year} {2021})}\BibitemShut
  {NoStop}%
\bibitem [{\citenamefont {Flower}\ \emph {et~al.}(2019)\citenamefont {Flower},
  \citenamefont {Bourhill}, \citenamefont {Goryachev},\ and\ \citenamefont
  {Tobar}}]{FLOWER2019100306}%
  \BibitemOpen
  \bibfield  {author} {\bibinfo {author} {\bibfnamefont {G.}~\bibnamefont
  {Flower}}, \bibinfo {author} {\bibfnamefont {J.}~\bibnamefont {Bourhill}},
  \bibinfo {author} {\bibfnamefont {M.}~\bibnamefont {Goryachev}},\ and\
  \bibinfo {author} {\bibfnamefont {M.~E.}\ \bibnamefont {Tobar}},\ }\bibfield
  {title} {\bibinfo {title} {Broadening frequency range of a ferromagnetic
  axion haloscope with strongly coupled cavity–magnon polaritons},\ }\href
  {https://doi.org/https://doi.org/10.1016/j.dark.2019.100306} {\bibfield
  {journal} {\bibinfo  {journal} {Phys. Dark Universe}\ }\textbf {\bibinfo
  {volume} {25}},\ \bibinfo {pages} {100306} (\bibinfo {year}
  {2019})}\BibitemShut {NoStop}%
\bibitem [{\citenamefont {Arvanitaki}\ and\ \citenamefont
  {Geraci}(2014)}]{AravanitakiPRL2014}%
  \BibitemOpen
  \bibfield  {author} {\bibinfo {author} {\bibfnamefont {A.}~\bibnamefont
  {Arvanitaki}}\ and\ \bibinfo {author} {\bibfnamefont {A.~A.}\ \bibnamefont
  {Geraci}},\ }\bibfield  {title} {\bibinfo {title} {Resonantly detecting
  axion-mediated forces with nuclear magnetic resonance},\ }\href
  {https://doi.org/10.1103/PhysRevLett.113.161801} {\bibfield  {journal}
  {\bibinfo  {journal} {Phys. Rev. Lett.}\ }\textbf {\bibinfo {volume} {113}},\
  \bibinfo {pages} {161801} (\bibinfo {year} {2014})}\BibitemShut {NoStop}%
\bibitem [{\citenamefont {Caldwell}\ \emph {et~al.}(2017)\citenamefont
  {Caldwell}, \citenamefont {Dvali}, \citenamefont {Majorovits}, \citenamefont
  {Millar}, \citenamefont {Raffelt}, \citenamefont {Redondo}, \citenamefont
  {Reimann}, \citenamefont {Simon},\ and\ \citenamefont
  {Steffen}}]{MadmaxPRL2017}%
  \BibitemOpen
  \bibfield  {author} {\bibinfo {author} {\bibfnamefont {A.}~\bibnamefont
  {Caldwell}}, \bibinfo {author} {\bibfnamefont {G.}~\bibnamefont {Dvali}},
  \bibinfo {author} {\bibfnamefont {B.}~\bibnamefont {Majorovits}}, \bibinfo
  {author} {\bibfnamefont {A.}~\bibnamefont {Millar}}, \bibinfo {author}
  {\bibfnamefont {G.}~\bibnamefont {Raffelt}}, \bibinfo {author} {\bibfnamefont
  {J.}~\bibnamefont {Redondo}}, \bibinfo {author} {\bibfnamefont
  {O.}~\bibnamefont {Reimann}}, \bibinfo {author} {\bibfnamefont
  {F.}~\bibnamefont {Simon}},\ and\ \bibinfo {author} {\bibfnamefont
  {F.}~\bibnamefont {Steffen}} (\bibinfo {collaboration} {MADMAX Working
  Group}),\ }\bibfield  {title} {\bibinfo {title} {Dielectric haloscopes: A new
  way to detect axion dark matter},\ }\href
  {https://doi.org/10.1103/PhysRevLett.118.091801} {\bibfield  {journal}
  {\bibinfo  {journal} {Phys. Rev. Lett.}\ }\textbf {\bibinfo {volume} {118}},\
  \bibinfo {pages} {091801} (\bibinfo {year} {2017})}\BibitemShut {NoStop}%
\bibitem [{\citenamefont {Barbieri}\ \emph {et~al.}(1989)\citenamefont
  {Barbieri}, \citenamefont {Cerdonio}, \citenamefont {Fiorentini},\ and\
  \citenamefont {Vitale}}]{BARBIERI1989357}%
  \BibitemOpen
  \bibfield  {author} {\bibinfo {author} {\bibfnamefont {R.}~\bibnamefont
  {Barbieri}}, \bibinfo {author} {\bibfnamefont {M.}~\bibnamefont {Cerdonio}},
  \bibinfo {author} {\bibfnamefont {G.}~\bibnamefont {Fiorentini}},\ and\
  \bibinfo {author} {\bibfnamefont {S.}~\bibnamefont {Vitale}},\ }\bibfield
  {title} {\bibinfo {title} {Axion to magnon conversion. a scheme for the
  detection of galactic axions},\ }\href
  {https://doi.org/https://doi.org/10.1016/0370-2693(89)91209-4} {\bibfield
  {journal} {\bibinfo  {journal} {Phys. Lett. B}\ }\textbf {\bibinfo {volume}
  {226}},\ \bibinfo {pages} {357} (\bibinfo {year} {1989})}\BibitemShut
  {NoStop}%
\bibitem [{\citenamefont {Crescini}\ \emph {et~al.}(2020)\citenamefont
  {Crescini}, \citenamefont {Alesini}, \citenamefont {Braggio}, \citenamefont
  {Carugno}, \citenamefont {D'Agostino}, \citenamefont {Di~Gioacchino},
  \citenamefont {Falferi}, \citenamefont {Gambardella}, \citenamefont {Gatti},
  \citenamefont {Iannone}, \citenamefont {Ligi}, \citenamefont {Lombardi},
  \citenamefont {Ortolan}, \citenamefont {Pengo}, \citenamefont {Ruoso},\ and\
  \citenamefont {Taffarello}}]{CresciniPRL2020}%
  \BibitemOpen
  \bibfield  {author} {\bibinfo {author} {\bibfnamefont {N.}~\bibnamefont
  {Crescini}}, \bibinfo {author} {\bibfnamefont {D.}~\bibnamefont {Alesini}},
  \bibinfo {author} {\bibfnamefont {C.}~\bibnamefont {Braggio}}, \bibinfo
  {author} {\bibfnamefont {G.}~\bibnamefont {Carugno}}, \bibinfo {author}
  {\bibfnamefont {D.}~\bibnamefont {D'Agostino}}, \bibinfo {author}
  {\bibfnamefont {D.}~\bibnamefont {Di~Gioacchino}}, \bibinfo {author}
  {\bibfnamefont {P.}~\bibnamefont {Falferi}}, \bibinfo {author} {\bibfnamefont
  {U.}~\bibnamefont {Gambardella}}, \bibinfo {author} {\bibfnamefont
  {C.}~\bibnamefont {Gatti}}, \bibinfo {author} {\bibfnamefont
  {G.}~\bibnamefont {Iannone}}, \bibinfo {author} {\bibfnamefont
  {C.}~\bibnamefont {Ligi}}, \bibinfo {author} {\bibfnamefont {A.}~\bibnamefont
  {Lombardi}}, \bibinfo {author} {\bibfnamefont {A.}~\bibnamefont {Ortolan}},
  \bibinfo {author} {\bibfnamefont {R.}~\bibnamefont {Pengo}}, \bibinfo
  {author} {\bibfnamefont {G.}~\bibnamefont {Ruoso}},\ and\ \bibinfo {author}
  {\bibfnamefont {L.}~\bibnamefont {Taffarello}} (\bibinfo {collaboration}
  {QUAX Collaboration}),\ }\bibfield  {title} {\bibinfo {title} {Axion search
  with a quantum-limited ferromagnetic haloscope},\ }\href
  {https://doi.org/10.1103/PhysRevLett.124.171801} {\bibfield  {journal}
  {\bibinfo  {journal} {Phys. Rev. Lett.}\ }\textbf {\bibinfo {volume} {124}},\
  \bibinfo {pages} {171801} (\bibinfo {year} {2020})}\BibitemShut {NoStop}%
\bibitem [{\citenamefont {Lamoreaux}\ \emph {et~al.}(2013)\citenamefont
  {Lamoreaux}, \citenamefont {van Bibber}, \citenamefont {Lehnert},\ and\
  \citenamefont {Carosi}}]{Lamoreaux2013}%
  \BibitemOpen
  \bibfield  {author} {\bibinfo {author} {\bibfnamefont {S.~K.}\ \bibnamefont
  {Lamoreaux}}, \bibinfo {author} {\bibfnamefont {K.~A.}\ \bibnamefont {van
  Bibber}}, \bibinfo {author} {\bibfnamefont {K.~W.}\ \bibnamefont {Lehnert}},\
  and\ \bibinfo {author} {\bibfnamefont {G.}~\bibnamefont {Carosi}},\
  }\bibfield  {title} {\bibinfo {title} {Analysis of single-photon and linear
  amplifier detectors for microwave cavity dark matter axion searches},\ }\href
  {https://doi.org/10.1103/PhysRevD.88.035020} {\bibfield  {journal} {\bibinfo
  {journal} {Phys. Rev. D}\ }\textbf {\bibinfo {volume} {88}},\ \bibinfo
  {pages} {035020} (\bibinfo {year} {2013})}\BibitemShut {NoStop}%
\bibitem [{\citenamefont {Dine}\ \emph {et~al.}(1981)\citenamefont {Dine},
  \citenamefont {Fischler},\ and\ \citenamefont {Srednicki}}]{DINE1981199}%
  \BibitemOpen
  \bibfield  {author} {\bibinfo {author} {\bibfnamefont {M.}~\bibnamefont
  {Dine}}, \bibinfo {author} {\bibfnamefont {W.}~\bibnamefont {Fischler}},\
  and\ \bibinfo {author} {\bibfnamefont {M.}~\bibnamefont {Srednicki}},\
  }\bibfield  {title} {\bibinfo {title} {A simple solution to the strong cp
  problem with a harmless axion},\ }\href
  {https://doi.org/https://doi.org/10.1016/0370-2693(81)90590-6} {\bibfield
  {journal} {\bibinfo  {journal} {Phys. Lett. B}\ }\textbf {\bibinfo {volume}
  {104}},\ \bibinfo {pages} {199} (\bibinfo {year} {1981})}\BibitemShut
  {NoStop}%
\bibitem [{\citenamefont {Zhitnitskii}(1980)}]{Zhitnitskii1980}%
  \BibitemOpen
  \bibfield  {author} {\bibinfo {author} {\bibfnamefont {A.~P.}\ \bibnamefont
  {Zhitnitskii}},\ }\bibfield  {title} {\bibinfo {title} {Possible suppression
  of axion-hadron interactions},\ }\href@noop {} {\bibfield  {journal}
  {\bibinfo  {journal} {Sov. J. Nucl. Phys.}\ }\textbf {\bibinfo {volume} {31}}
  (\bibinfo {year} {1980})}\BibitemShut {NoStop}%
\bibitem [{\citenamefont {Knirck}\ \emph {et~al.}(2018)\citenamefont {Knirck},
  \citenamefont {Millar}, \citenamefont {O{\textquotesingle}Hare},
  \citenamefont {Redondo},\ and\ \citenamefont {Steffen}}]{Knirck_2018}%
  \BibitemOpen
  \bibfield  {author} {\bibinfo {author} {\bibfnamefont {S.}~\bibnamefont
  {Knirck}}, \bibinfo {author} {\bibfnamefont {A.~J.}\ \bibnamefont {Millar}},
  \bibinfo {author} {\bibfnamefont {C.~A.}\ \bibnamefont
  {O{\textquotesingle}Hare}}, \bibinfo {author} {\bibfnamefont
  {J.}~\bibnamefont {Redondo}},\ and\ \bibinfo {author} {\bibfnamefont {F.~D.}\
  \bibnamefont {Steffen}},\ }\bibfield  {title} {\bibinfo {title} {Directional
  axion detection},\ }\href {https://doi.org/10.1088/1475-7516/2018/11/051}
  {\bibfield  {journal} {\bibinfo  {journal} {J. Cosmol. Astropart. Phys.}\
  }\textbf {\bibinfo {volume} {2018}}\bibinfo  {number} { (11)},\ \bibinfo
  {pages} {051}}\BibitemShut {NoStop}%
\bibitem [{\citenamefont {Kittel}(2005)}]{KittelSolidState}%
  \BibitemOpen
\bibfield  {number} {  }\bibfield  {author} {\bibinfo {author} {\bibfnamefont
  {C.}~\bibnamefont {Kittel}},\ }\href@noop {} {\emph {\bibinfo {title}
  {Introduction to solid state physics}}},\ \bibinfo {edition} {8th}\ ed.\
  (\bibinfo  {publisher} {Wiley},\ \bibinfo {address} {Hoboken, N.J},\ \bibinfo
  {year} {2005})\BibitemShut {NoStop}%
\bibitem [{\citenamefont {Deeter}\ \emph {et~al.}(1990)\citenamefont {Deeter},
  \citenamefont {Rose},\ and\ \citenamefont {Day}}]{Deeter1990}%
  \BibitemOpen
  \bibfield  {author} {\bibinfo {author} {\bibfnamefont {M.}~\bibnamefont
  {Deeter}}, \bibinfo {author} {\bibfnamefont {A.}~\bibnamefont {Rose}},\ and\
  \bibinfo {author} {\bibfnamefont {G.}~\bibnamefont {Day}},\ }\bibfield
  {title} {\bibinfo {title} {Fast, sensitive magnetic-field sensors based on
  the faraday effect in {YIG}},\ }\href {https://doi.org/10.1109/50.62880}
  {\bibfield  {journal} {\bibinfo  {journal} {J. Light. Technol.}\ }\textbf
  {\bibinfo {volume} {8}},\ \bibinfo {pages} {1838} (\bibinfo {year}
  {1990})}\BibitemShut {NoStop}%
\bibitem [{\citenamefont {Hisatomi}\ \emph {et~al.}(2016)\citenamefont
  {Hisatomi}, \citenamefont {Osada}, \citenamefont {Tabuchi}, \citenamefont
  {Ishikawa}, \citenamefont {Noguchi}, \citenamefont {Yamazaki}, \citenamefont
  {Usami},\ and\ \citenamefont {Nakamura}}]{HisatomiPRB2016}%
  \BibitemOpen
  \bibfield  {author} {\bibinfo {author} {\bibfnamefont {R.}~\bibnamefont
  {Hisatomi}}, \bibinfo {author} {\bibfnamefont {A.}~\bibnamefont {Osada}},
  \bibinfo {author} {\bibfnamefont {Y.}~\bibnamefont {Tabuchi}}, \bibinfo
  {author} {\bibfnamefont {T.}~\bibnamefont {Ishikawa}}, \bibinfo {author}
  {\bibfnamefont {A.}~\bibnamefont {Noguchi}}, \bibinfo {author} {\bibfnamefont
  {R.}~\bibnamefont {Yamazaki}}, \bibinfo {author} {\bibfnamefont
  {K.}~\bibnamefont {Usami}},\ and\ \bibinfo {author} {\bibfnamefont
  {Y.}~\bibnamefont {Nakamura}},\ }\bibfield  {title} {\bibinfo {title}
  {Bidirectional conversion between microwave and light via ferromagnetic
  magnons},\ }\href {https://doi.org/10.1103/PhysRevB.93.174427} {\bibfield
  {journal} {\bibinfo  {journal} {Phys. Rev. B}\ }\textbf {\bibinfo {volume}
  {93}},\ \bibinfo {pages} {174427} (\bibinfo {year} {2016})}\BibitemShut
  {NoStop}%
\bibitem [{\citenamefont {Dillon}\ \emph {et~al.}(1963)\citenamefont {Dillon},
  \citenamefont {Kamimura},\ and\ \citenamefont {Remeika}}]{DillonJAP1963}%
  \BibitemOpen
  \bibfield  {author} {\bibinfo {author} {\bibfnamefont {J.~F.}\ \bibnamefont
  {Dillon}}, \bibinfo {author} {\bibfnamefont {H.}~\bibnamefont {Kamimura}},\
  and\ \bibinfo {author} {\bibfnamefont {J.~P.}\ \bibnamefont {Remeika}},\
  }\bibfield  {title} {\bibinfo {title} {Magneto‐optical studies of chromium
  tribromide},\ }\href {https://doi.org/10.1063/1.1729455} {\bibfield
  {journal} {\bibinfo  {journal} {J. Appl. Phys.}\ }\textbf {\bibinfo {volume}
  {34}},\ \bibinfo {pages} {1240} (\bibinfo {year} {1963})}\BibitemShut
  {NoStop}%
\bibitem [{\citenamefont {Hemmati}\ \emph {et~al.}(2019)\citenamefont
  {Hemmati}, \citenamefont {Bootpakdeetam},\ and\ \citenamefont
  {Magnusson}}]{Hemmati:19}%
  \BibitemOpen
  \bibfield  {author} {\bibinfo {author} {\bibfnamefont {H.}~\bibnamefont
  {Hemmati}}, \bibinfo {author} {\bibfnamefont {P.}~\bibnamefont
  {Bootpakdeetam}},\ and\ \bibinfo {author} {\bibfnamefont {R.}~\bibnamefont
  {Magnusson}},\ }\bibfield  {title} {\bibinfo {title} {Metamaterial polarizer
  providing principally unlimited extinction},\ }\href
  {https://doi.org/10.1364/OL.44.005630} {\bibfield  {journal} {\bibinfo
  {journal} {Opt. Lett.}\ }\textbf {\bibinfo {volume} {44}},\ \bibinfo {pages}
  {5630} (\bibinfo {year} {2019})}\BibitemShut {NoStop}%
\bibitem [{\citenamefont {den Boer}\ \emph {et~al.}(1995)\citenamefont {den
  Boer}, \citenamefont {Kroesen}, \citenamefont {de~Zeeuw},\ and\ \citenamefont
  {de~Hoog}}]{denBoer:95}%
  \BibitemOpen
  \bibfield  {author} {\bibinfo {author} {\bibfnamefont {J.~H. W.~G.}\
  \bibnamefont {den Boer}}, \bibinfo {author} {\bibfnamefont {G.~M.~W.}\
  \bibnamefont {Kroesen}}, \bibinfo {author} {\bibfnamefont {W.}~\bibnamefont
  {de~Zeeuw}},\ and\ \bibinfo {author} {\bibfnamefont {F.~J.}\ \bibnamefont
  {de~Hoog}},\ }\bibfield  {title} {\bibinfo {title} {Improved polarizer in the
  infrared: two wire-grid polarizers in tandem},\ }\href
  {https://doi.org/10.1364/OL.20.000800} {\bibfield  {journal} {\bibinfo
  {journal} {Opt. Lett.}\ }\textbf {\bibinfo {volume} {20}},\ \bibinfo {pages}
  {800} (\bibinfo {year} {1995})}\BibitemShut {NoStop}%
\bibitem [{\citenamefont {Fox}(2006)}]{QuantumOpticsFox}%
  \BibitemOpen
  \bibfield  {author} {\bibinfo {author} {\bibfnamefont {M.}~\bibnamefont
  {Fox}},\ }\href@noop {} {\emph {\bibinfo {title} {Quantum Optics: An
  Introduction}}}\ (\bibinfo  {publisher} {Oxford University Press},\ \bibinfo
  {year} {2006})\BibitemShut {NoStop}%
\bibitem [{\citenamefont {Ceccarelli}\ \emph {et~al.}(2021)\citenamefont
  {Ceccarelli}, \citenamefont {Acconcia}, \citenamefont {Gulinatti},
  \citenamefont {Ghioni}, \citenamefont {Rech},\ and\ \citenamefont
  {Osellame}}]{Ceccarelli2021}%
  \BibitemOpen
  \bibfield  {author} {\bibinfo {author} {\bibfnamefont {F.}~\bibnamefont
  {Ceccarelli}}, \bibinfo {author} {\bibfnamefont {G.}~\bibnamefont
  {Acconcia}}, \bibinfo {author} {\bibfnamefont {A.}~\bibnamefont {Gulinatti}},
  \bibinfo {author} {\bibfnamefont {M.}~\bibnamefont {Ghioni}}, \bibinfo
  {author} {\bibfnamefont {I.}~\bibnamefont {Rech}},\ and\ \bibinfo {author}
  {\bibfnamefont {R.}~\bibnamefont {Osellame}},\ }\bibfield  {title} {\bibinfo
  {title} {Recent advances and future perspectives of single-photon avalanche
  diodes for quantum photonics applications},\ }\href
  {https://doi.org/https://doi.org/10.1002/qute.202000102} {\bibfield
  {journal} {\bibinfo  {journal} {Adv. Quantum Technol.}\ }\textbf {\bibinfo
  {volume} {4}},\ \bibinfo {pages} {2000102} (\bibinfo {year}
  {2021})}\BibitemShut {NoStop}%
\bibitem [{\citenamefont {Esmaeil~Zadeh}\ \emph {et~al.}(2021)\citenamefont
  {Esmaeil~Zadeh}, \citenamefont {Chang}, \citenamefont {Los}, \citenamefont
  {Gyger}, \citenamefont {Elshaari}, \citenamefont {Steinhauer}, \citenamefont
  {Dorenbos},\ and\ \citenamefont {Zwiller}}]{ZwillerAPL2021}%
  \BibitemOpen
  \bibfield  {author} {\bibinfo {author} {\bibfnamefont {I.}~\bibnamefont
  {Esmaeil~Zadeh}}, \bibinfo {author} {\bibfnamefont {J.}~\bibnamefont
  {Chang}}, \bibinfo {author} {\bibfnamefont {J.~W.~N.}\ \bibnamefont {Los}},
  \bibinfo {author} {\bibfnamefont {S.}~\bibnamefont {Gyger}}, \bibinfo
  {author} {\bibfnamefont {A.~W.}\ \bibnamefont {Elshaari}}, \bibinfo {author}
  {\bibfnamefont {S.}~\bibnamefont {Steinhauer}}, \bibinfo {author}
  {\bibfnamefont {S.~N.}\ \bibnamefont {Dorenbos}},\ and\ \bibinfo {author}
  {\bibfnamefont {V.}~\bibnamefont {Zwiller}},\ }\bibfield  {title} {\bibinfo
  {title} {Superconducting nanowire single-photon detectors: A perspective on
  evolution, state-of-the-art, future developments, and applications},\ }\href
  {https://doi.org/10.1063/5.0045990} {\bibfield  {journal} {\bibinfo
  {journal} {Appl. Phys. Lett.}\ }\textbf {\bibinfo {volume} {118}},\ \bibinfo
  {pages} {190502} (\bibinfo {year} {2021})}\BibitemShut {NoStop}%
\bibitem [{\citenamefont {Schweickert}\ \emph {et~al.}(2018)\citenamefont
  {Schweickert}, \citenamefont {Jöns}, \citenamefont {Zeuner}, \citenamefont
  {Covre~da Silva}, \citenamefont {Huang}, \citenamefont {Lettner},
  \citenamefont {Reindl}, \citenamefont {Zichi}, \citenamefont {Trotta},
  \citenamefont {Rastelli},\ and\ \citenamefont
  {Zwiller}}]{SchweickertAPL2018}%
  \BibitemOpen
  \bibfield  {author} {\bibinfo {author} {\bibfnamefont {L.}~\bibnamefont
  {Schweickert}}, \bibinfo {author} {\bibfnamefont {K.~D.}\ \bibnamefont
  {Jöns}}, \bibinfo {author} {\bibfnamefont {K.~D.}\ \bibnamefont {Zeuner}},
  \bibinfo {author} {\bibfnamefont {S.~F.}\ \bibnamefont {Covre~da Silva}},
  \bibinfo {author} {\bibfnamefont {H.}~\bibnamefont {Huang}}, \bibinfo
  {author} {\bibfnamefont {T.}~\bibnamefont {Lettner}}, \bibinfo {author}
  {\bibfnamefont {M.}~\bibnamefont {Reindl}}, \bibinfo {author} {\bibfnamefont
  {J.}~\bibnamefont {Zichi}}, \bibinfo {author} {\bibfnamefont
  {R.}~\bibnamefont {Trotta}}, \bibinfo {author} {\bibfnamefont
  {A.}~\bibnamefont {Rastelli}},\ and\ \bibinfo {author} {\bibfnamefont
  {V.}~\bibnamefont {Zwiller}},\ }\bibfield  {title} {\bibinfo {title}
  {On-demand generation of background-free single photons from a solid-state
  source},\ }\href {https://doi.org/10.1063/1.5020038} {\bibfield  {journal}
  {\bibinfo  {journal} {Appl. Phys. Lett.}\ }\textbf {\bibinfo {volume}
  {112}},\ \bibinfo {pages} {093106} (\bibinfo {year} {2018})}\BibitemShut
  {NoStop}%
\bibitem [{\citenamefont {Lage}\ \emph {et~al.}(2017)\citenamefont {Lage},
  \citenamefont {Beran}, \citenamefont {Quindeau}, \citenamefont {Ohnoutek},
  \citenamefont {Kucera}, \citenamefont {Antos}, \citenamefont {Sani},
  \citenamefont {Dionne}, \citenamefont {Veis},\ and\ \citenamefont
  {Ross}}]{LageAPL2017}%
  \BibitemOpen
  \bibfield  {author} {\bibinfo {author} {\bibfnamefont {E.}~\bibnamefont
  {Lage}}, \bibinfo {author} {\bibfnamefont {L.}~\bibnamefont {Beran}},
  \bibinfo {author} {\bibfnamefont {A.~U.}\ \bibnamefont {Quindeau}}, \bibinfo
  {author} {\bibfnamefont {L.}~\bibnamefont {Ohnoutek}}, \bibinfo {author}
  {\bibfnamefont {M.}~\bibnamefont {Kucera}}, \bibinfo {author} {\bibfnamefont
  {R.}~\bibnamefont {Antos}}, \bibinfo {author} {\bibfnamefont {S.~R.}\
  \bibnamefont {Sani}}, \bibinfo {author} {\bibfnamefont {G.~F.}\ \bibnamefont
  {Dionne}}, \bibinfo {author} {\bibfnamefont {M.}~\bibnamefont {Veis}},\ and\
  \bibinfo {author} {\bibfnamefont {C.~A.}\ \bibnamefont {Ross}},\ }\bibfield
  {title} {\bibinfo {title} {Temperature-dependent faraday rotation and
  magnetization reorientation in cerium-substituted yttrium iron garnet thin
  films},\ }\href {https://doi.org/10.1063/1.4976817} {\bibfield  {journal}
  {\bibinfo  {journal} {APL Mater.}\ }\textbf {\bibinfo {volume} {5}},\
  \bibinfo {pages} {036104} (\bibinfo {year} {2017})}\BibitemShut {NoStop}%
\bibitem [{\citenamefont {Higuchi}\ \emph {et~al.}(2003)\citenamefont
  {Higuchi}, \citenamefont {Furukawa}, \citenamefont {Takekawa}, \citenamefont
  {Kamada}, \citenamefont {Kitamura},\ and\ \citenamefont
  {Uyeda}}]{HIGUCHI2003293}%
  \BibitemOpen
  \bibfield  {author} {\bibinfo {author} {\bibfnamefont {S.}~\bibnamefont
  {Higuchi}}, \bibinfo {author} {\bibfnamefont {Y.}~\bibnamefont {Furukawa}},
  \bibinfo {author} {\bibfnamefont {S.}~\bibnamefont {Takekawa}}, \bibinfo
  {author} {\bibfnamefont {O.}~\bibnamefont {Kamada}}, \bibinfo {author}
  {\bibfnamefont {K.}~\bibnamefont {Kitamura}},\ and\ \bibinfo {author}
  {\bibfnamefont {K.}~\bibnamefont {Uyeda}},\ }\bibfield  {title} {\bibinfo
  {title} {Magnetooptical properties of cerium-substituted yttrium iron garnet
  single crystals for magnetic-field sensor},\ }\href
  {https://doi.org/https://doi.org/10.1016/S0924-4247(03)00104-3} {\bibfield
  {journal} {\bibinfo  {journal} {Sens. Actuator A Phys.}\ }\textbf {\bibinfo
  {volume} {105}},\ \bibinfo {pages} {293} (\bibinfo {year}
  {2003})}\BibitemShut {NoStop}%
\bibitem [{\citenamefont {Lachance-Quirion}\ \emph {et~al.}(2019)\citenamefont
  {Lachance-Quirion}, \citenamefont {Tabuchi}, \citenamefont {Gloppe},
  \citenamefont {Usami},\ and\ \citenamefont
  {Nakamura}}]{Lachance_Quirion_2019}%
  \BibitemOpen
  \bibfield  {author} {\bibinfo {author} {\bibfnamefont {D.}~\bibnamefont
  {Lachance-Quirion}}, \bibinfo {author} {\bibfnamefont {Y.}~\bibnamefont
  {Tabuchi}}, \bibinfo {author} {\bibfnamefont {A.}~\bibnamefont {Gloppe}},
  \bibinfo {author} {\bibfnamefont {K.}~\bibnamefont {Usami}},\ and\ \bibinfo
  {author} {\bibfnamefont {Y.}~\bibnamefont {Nakamura}},\ }\bibfield  {title}
  {\bibinfo {title} {Hybrid quantum systems based on magnonics},\ }\href@noop
  {} {\bibfield  {journal} {\bibinfo  {journal} {Appl. Phys. Express.}\
  }\textbf {\bibinfo {volume} {12}},\ \bibinfo {pages} {070101} (\bibinfo
  {year} {2019})}\BibitemShut {NoStop}%
\bibitem [{\citenamefont {Rahman}(2019)}]{Rahman_2019}%
  \BibitemOpen
  \bibfield  {author} {\bibinfo {author} {\bibfnamefont {A.~T. M.~A.}\
  \bibnamefont {Rahman}},\ }\bibfield  {title} {\bibinfo {title} {Large spatial
  schrödinger cat state using a levitated ferrimagnetic nanoparticle},\ }\href
  {https://doi.org/10.1088/1367-2630/ab4fb3} {\bibfield  {journal} {\bibinfo
  {journal} {New J. Phys.}\ }\textbf {\bibinfo {volume} {21}},\ \bibinfo
  {pages} {113011} (\bibinfo {year} {2019})}\BibitemShut {NoStop}%
\bibitem [{\citenamefont {Higuchi}\ \emph {et~al.}(1999)\citenamefont
  {Higuchi}, \citenamefont {Furukawa}, \citenamefont {Takekawa}, \citenamefont
  {Kamada},\ and\ \citenamefont {Kitamura}}]{Higuchi_1999}%
  \BibitemOpen
  \bibfield  {author} {\bibinfo {author} {\bibfnamefont {S.}~\bibnamefont
  {Higuchi}}, \bibinfo {author} {\bibfnamefont {Y.}~\bibnamefont {Furukawa}},
  \bibinfo {author} {\bibfnamefont {S.}~\bibnamefont {Takekawa}}, \bibinfo
  {author} {\bibfnamefont {O.}~\bibnamefont {Kamada}},\ and\ \bibinfo {author}
  {\bibfnamefont {K.}~\bibnamefont {Kitamura}},\ }\bibfield  {title} {\bibinfo
  {title} {Magneto-optical properties of cerium-substituted yttrium iron garnet
  single crystals grown by traveling solvent floating zone method},\ }\href
  {https://doi.org/10.1143/jjap.38.4122} {\bibfield  {journal} {\bibinfo
  {journal} {Jpn. J. Appl. Phys.}\ }\textbf {\bibinfo {volume} {38}},\ \bibinfo
  {pages} {4122} (\bibinfo {year} {1999})}\BibitemShut {NoStop}%
\bibitem [{Rio()}]{RioGrande1550nm}%
  \BibitemOpen
  \bibfield  {title} {\bibinfo {title} {Rio {GRANDE}},\ }\href@noop {}
  {\bibinfo  {journal} {https://rio-lasers.com/}\ }\BibitemShut {NoStop}%
\bibitem [{\citenamefont {Meylahn}\ \emph {et~al.}(2022)\citenamefont
  {Meylahn}, \citenamefont {Knust},\ and\ \citenamefont
  {Willke}}]{MeylahnPRD2022}%
  \BibitemOpen
\bibfield  {journal} {  }\bibfield  {author} {\bibinfo {author} {\bibfnamefont
  {F.}~\bibnamefont {Meylahn}}, \bibinfo {author} {\bibfnamefont
  {N.}~\bibnamefont {Knust}},\ and\ \bibinfo {author} {\bibfnamefont
  {B.}~\bibnamefont {Willke}},\ }\bibfield  {title} {\bibinfo {title}
  {Stabilized laser system at 1550 nm wavelength for future gravitational-wave
  detectors},\ }\href {https://doi.org/10.1103/PhysRevD.105.122004} {\bibfield
  {journal} {\bibinfo  {journal} {Phys. Rev. D}\ }\textbf {\bibinfo {volume}
  {105}},\ \bibinfo {pages} {122004} (\bibinfo {year} {2022})}\BibitemShut
  {NoStop}%
\bibitem [{\citenamefont {Callon}\ \emph {et~al.}(2021)\citenamefont {Callon},
  \citenamefont {Malär}, \citenamefont {Pfister}, \citenamefont {Římal},
  \citenamefont {Weber}, \citenamefont {Wiegand}, \citenamefont {Zehnder},
  \citenamefont {Chávez}, \citenamefont {Cadalbert}, \citenamefont {Deb},
  \citenamefont {Däpp}, \citenamefont {Fogeron}, \citenamefont {Hunkeler},
  \citenamefont {Lecoq}, \citenamefont {Torosyan}, \citenamefont {Zyla},
  \citenamefont {Glockshuber}, \citenamefont {Jonas}, \citenamefont {Nassal},
  \citenamefont {Ernst}, \citenamefont {Böckmann},\ and\ \citenamefont
  {Meier}}]{CallonMorgane2021}%
  \BibitemOpen
  \bibfield  {author} {\bibinfo {author} {\bibfnamefont {M.}~\bibnamefont
  {Callon}}, \bibinfo {author} {\bibfnamefont {A.~A.}\ \bibnamefont {Malär}},
  \bibinfo {author} {\bibfnamefont {S.}~\bibnamefont {Pfister}}, \bibinfo
  {author} {\bibfnamefont {V.}~\bibnamefont {Římal}}, \bibinfo {author}
  {\bibfnamefont {M.~E.}\ \bibnamefont {Weber}}, \bibinfo {author}
  {\bibfnamefont {T.}~\bibnamefont {Wiegand}}, \bibinfo {author} {\bibfnamefont
  {J.}~\bibnamefont {Zehnder}}, \bibinfo {author} {\bibfnamefont
  {M.}~\bibnamefont {Chávez}}, \bibinfo {author} {\bibfnamefont
  {R.}~\bibnamefont {Cadalbert}}, \bibinfo {author} {\bibfnamefont
  {R.}~\bibnamefont {Deb}}, \bibinfo {author} {\bibfnamefont {A.}~\bibnamefont
  {Däpp}}, \bibinfo {author} {\bibfnamefont {M.-L.}\ \bibnamefont {Fogeron}},
  \bibinfo {author} {\bibfnamefont {A.}~\bibnamefont {Hunkeler}}, \bibinfo
  {author} {\bibfnamefont {L.}~\bibnamefont {Lecoq}}, \bibinfo {author}
  {\bibfnamefont {A.}~\bibnamefont {Torosyan}}, \bibinfo {author}
  {\bibfnamefont {D.}~\bibnamefont {Zyla}}, \bibinfo {author} {\bibfnamefont
  {R.}~\bibnamefont {Glockshuber}}, \bibinfo {author} {\bibfnamefont
  {S.}~\bibnamefont {Jonas}}, \bibinfo {author} {\bibfnamefont
  {M.}~\bibnamefont {Nassal}}, \bibinfo {author} {\bibfnamefont
  {M.}~\bibnamefont {Ernst}}, \bibinfo {author} {\bibfnamefont
  {A.}~\bibnamefont {Böckmann}},\ and\ \bibinfo {author} {\bibfnamefont
  {B.~H.}\ \bibnamefont {Meier}},\ }\bibfield  {title} {\bibinfo {title}
  {Biomolecular solid-state nmr spectroscopy at 1200 mhz: the gain in
  resolution},\ }\href@noop {} {\bibfield  {journal} {\bibinfo  {journal} {J.
  Biomol. NMR}\ }\textbf {\bibinfo {volume} {75}},\ \bibinfo {pages} {255}
  (\bibinfo {year} {2021})}\BibitemShut {NoStop}%
\bibitem [{\citenamefont {Hahn}\ \emph {et~al.}(2019)\citenamefont {Hahn},
  \citenamefont {Kim}, \citenamefont {Kim}, \citenamefont {Hu}, \citenamefont
  {Painter}, \citenamefont {Dixon}, \citenamefont {Kim}, \citenamefont
  {Bhattarai}, \citenamefont {Noguchi}, \citenamefont {Jaroszynski},\ and\
  \citenamefont {Larbalestier}}]{HahnSeungyong2019}%
  \BibitemOpen
  \bibfield  {author} {\bibinfo {author} {\bibfnamefont {S.}~\bibnamefont
  {Hahn}}, \bibinfo {author} {\bibfnamefont {K.}~\bibnamefont {Kim}}, \bibinfo
  {author} {\bibfnamefont {K.}~\bibnamefont {Kim}}, \bibinfo {author}
  {\bibfnamefont {X.}~\bibnamefont {Hu}}, \bibinfo {author} {\bibfnamefont
  {T.}~\bibnamefont {Painter}}, \bibinfo {author} {\bibfnamefont
  {I.}~\bibnamefont {Dixon}}, \bibinfo {author} {\bibfnamefont
  {S.}~\bibnamefont {Kim}}, \bibinfo {author} {\bibfnamefont {K.~R.}\
  \bibnamefont {Bhattarai}}, \bibinfo {author} {\bibfnamefont {S.}~\bibnamefont
  {Noguchi}}, \bibinfo {author} {\bibfnamefont {J.}~\bibnamefont
  {Jaroszynski}},\ and\ \bibinfo {author} {\bibfnamefont {D.~C.}\ \bibnamefont
  {Larbalestier}},\ }\bibfield  {title} {\bibinfo {title} {45.5-tesla
  direct-current magnetic field generated with a high-temperature
  superconducting magnet},\ }\href@noop {} {\bibfield  {journal} {\bibinfo
  {journal} {Nature}\ }\textbf {\bibinfo {volume} {570}},\ \bibinfo {pages}
  {496} (\bibinfo {year} {2019})}\BibitemShut {NoStop}%
\bibitem [{\citenamefont {Miller}(2003)}]{MillerIEEE2003}%
  \BibitemOpen
  \bibfield  {author} {\bibinfo {author} {\bibfnamefont {J.}~\bibnamefont
  {Miller}},\ }\bibfield  {title} {\bibinfo {title} {The {NHMFL} {45-T} hybrid
  magnet system: past, present, and future},\ }\href
  {https://doi.org/10.1109/TASC.2003.812673} {\bibfield  {journal} {\bibinfo
  {journal} {IEEE Trans. Appl. Supercond}\ }\textbf {\bibinfo {volume} {13}},\
  \bibinfo {pages} {1385} (\bibinfo {year} {2003})}\BibitemShut {NoStop}%
\bibitem [{\citenamefont {Rempe}\ \emph {et~al.}(1992)\citenamefont {Rempe},
  \citenamefont {Thompson}, \citenamefont {Kimble},\ and\ \citenamefont
  {Lalezari}}]{RempeOL1992}%
  \BibitemOpen
  \bibfield  {author} {\bibinfo {author} {\bibfnamefont {G.}~\bibnamefont
  {Rempe}}, \bibinfo {author} {\bibfnamefont {R.~J.}\ \bibnamefont {Thompson}},
  \bibinfo {author} {\bibfnamefont {H.~J.}\ \bibnamefont {Kimble}},\ and\
  \bibinfo {author} {\bibfnamefont {R.}~\bibnamefont {Lalezari}},\ }\bibfield
  {title} {\bibinfo {title} {Measurement of ultralow losses in an optical
  interferometer},\ }\href {https://doi.org/10.1364/OL.17.000363} {\bibfield
  {journal} {\bibinfo  {journal} {Opt. Lett.}\ }\textbf {\bibinfo {volume}
  {17}},\ \bibinfo {pages} {363} (\bibinfo {year} {1992})}\BibitemShut
  {NoStop}%
\bibitem [{\citenamefont {Spencer}\ \emph {et~al.}(1961)\citenamefont
  {Spencer}, \citenamefont {LeCraw},\ and\ \citenamefont
  {Linares}}]{Spencer1961}%
  \BibitemOpen
  \bibfield  {author} {\bibinfo {author} {\bibfnamefont {E.~G.}\ \bibnamefont
  {Spencer}}, \bibinfo {author} {\bibfnamefont {R.~C.}\ \bibnamefont
  {LeCraw}},\ and\ \bibinfo {author} {\bibfnamefont {R.~C.}\ \bibnamefont
  {Linares}},\ }\bibfield  {title} {\bibinfo {title} {Low-temperature
  ferromagnetic relaxation in yttrium iron garnet},\ }\href
  {https://doi.org/10.1103/PhysRev.123.1937} {\bibfield  {journal} {\bibinfo
  {journal} {Phys. Rev.}\ }\textbf {\bibinfo {volume} {123}},\ \bibinfo {pages}
  {1937} (\bibinfo {year} {1961})}\BibitemShut {NoStop}%
\bibitem [{\citenamefont {Aasi}\ \emph {et~al.}(2013)\citenamefont {Aasi},
  \citenamefont {Sluys}, \citenamefont {Zweizig},\ and\ \citenamefont
  {et~al}}]{AasiJ2013Esot}%
  \BibitemOpen
  \bibfield  {author} {\bibinfo {author} {\bibfnamefont {J.}~\bibnamefont
  {Aasi}}, \bibinfo {author} {\bibfnamefont {M.~v.~d.}\ \bibnamefont {Sluys}},
  \bibinfo {author} {\bibfnamefont {J.}~\bibnamefont {Zweizig}},\ and\ \bibinfo
  {author} {\bibnamefont {et~al}},\ }\bibfield  {title} {\bibinfo {title}
  {Enhanced sensitivity of the ligo gravitational wave detector by using
  squeezed states of light},\ }\href@noop {} {\bibfield  {journal} {\bibinfo
  {journal} {Nat. Photon.}\ }\textbf {\bibinfo {volume} {7}},\ \bibinfo {pages}
  {613} (\bibinfo {year} {2013})}\BibitemShut {NoStop}%
\bibitem [{\citenamefont {Hisatomi}\ \emph {et~al.}(2019)\citenamefont
  {Hisatomi}, \citenamefont {Noguchi}, \citenamefont {Yamazaki}, \citenamefont
  {Nakata}, \citenamefont {Gloppe}, \citenamefont {Nakamura},\ and\
  \citenamefont {Usami}}]{HisatomiPRL2019}%
  \BibitemOpen
  \bibfield  {author} {\bibinfo {author} {\bibfnamefont {R.}~\bibnamefont
  {Hisatomi}}, \bibinfo {author} {\bibfnamefont {A.}~\bibnamefont {Noguchi}},
  \bibinfo {author} {\bibfnamefont {R.}~\bibnamefont {Yamazaki}}, \bibinfo
  {author} {\bibfnamefont {Y.}~\bibnamefont {Nakata}}, \bibinfo {author}
  {\bibfnamefont {A.}~\bibnamefont {Gloppe}}, \bibinfo {author} {\bibfnamefont
  {Y.}~\bibnamefont {Nakamura}},\ and\ \bibinfo {author} {\bibfnamefont
  {K.}~\bibnamefont {Usami}},\ }\bibfield  {title} {\bibinfo {title}
  {Helicity-changing brillouin light scattering by magnons in a ferromagnetic
  crystal},\ }\href {https://doi.org/10.1103/PhysRevLett.123.207401} {\bibfield
   {journal} {\bibinfo  {journal} {Phys. Rev. Lett.}\ }\textbf {\bibinfo
  {volume} {123}},\ \bibinfo {pages} {207401} (\bibinfo {year}
  {2019})}\BibitemShut {NoStop}%
\bibitem [{\citenamefont {Marsh}\ \emph {et~al.}(2019)\citenamefont {Marsh},
  \citenamefont {Fong}, \citenamefont {Lentz}, \citenamefont
  {\ifmmode~\check{S}\else \v{S}\fi{}mejkal},\ and\ \citenamefont
  {Ali}}]{MarshPRL2019}%
  \BibitemOpen
  \bibfield  {author} {\bibinfo {author} {\bibfnamefont {D.~J.~E.}\
  \bibnamefont {Marsh}}, \bibinfo {author} {\bibfnamefont {K.~C.}\ \bibnamefont
  {Fong}}, \bibinfo {author} {\bibfnamefont {E.~W.}\ \bibnamefont {Lentz}},
  \bibinfo {author} {\bibfnamefont {L.}~\bibnamefont {\ifmmode~\check{S}\else
  \v{S}\fi{}mejkal}},\ and\ \bibinfo {author} {\bibfnamefont {M.~N.}\
  \bibnamefont {Ali}},\ }\bibfield  {title} {\bibinfo {title} {Proposal to
  detect dark matter using axionic topological antiferromagnets},\ }\href
  {https://doi.org/10.1103/PhysRevLett.123.121601} {\bibfield  {journal}
  {\bibinfo  {journal} {Phys. Rev. Lett.}\ }\textbf {\bibinfo {volume} {123}},\
  \bibinfo {pages} {121601} (\bibinfo {year} {2019})}\BibitemShut {NoStop}%
\end{thebibliography}


\end{document}